\documentclass[prx, aps,
 twocolumn,
 superscriptaddress,
 amsmath,amssymb,10pt,
]{revtex4-2}

\usepackage{times}
\usepackage{color}
\usepackage{graphicx}
\usepackage[dvipsnames]{xcolor}
\usepackage{physics}
\usepackage{bm}
\usepackage{mathtools}
\usepackage{upgreek}
\usepackage{footmisc}
\usepackage{makecell}
\usepackage{soul}
\usepackage{lipsum}
\usepackage{wasysym}
\usepackage[colorlinks=true,allcolors=blue]{hyperref}

\begin{document}

\title{Floquet-Engineered Fast SNAP Gates in Weakly Coupled Circuit-QED Systems}

\author{Xinyuan You}
\email[xinyuan@fnal.gov]{}
\affiliation{Superconducting Quantum Materials and Systems Division, Fermi National Accelerator Laboratory (FNAL), Batavia, IL 60510, USA}

\author{Andy C.~Y.~Li}
\affiliation{Superconducting Quantum Materials and Systems Division, Fermi National Accelerator Laboratory (FNAL), Batavia, IL 60510, USA}

\author{Tanay Roy}
\affiliation{Superconducting Quantum Materials and Systems Division, Fermi National Accelerator Laboratory (FNAL), Batavia, IL 60510, USA}

\author{Shaojiang Zhu}
\affiliation{Superconducting Quantum Materials and Systems Division, Fermi National Accelerator Laboratory (FNAL), Batavia, IL 60510, USA} 

\author{Alexander Romanenko}
\affiliation{Superconducting Quantum Materials and Systems Division, Fermi National Accelerator Laboratory (FNAL), Batavia, IL 60510, USA}

\author{\\Anna Grassellino}
\affiliation{Superconducting Quantum Materials and Systems Division, Fermi National Accelerator Laboratory (FNAL), Batavia, IL 60510, USA}

\author{Yao Lu}
\email[yaolu@fnal.gov]{}
\affiliation{Superconducting Quantum Materials and Systems Division, Fermi National Accelerator Laboratory (FNAL), Batavia, IL 60510, USA}

\author{Srivatsan Chakram}
\email[schakram@physics.rutgers.edu]{}
\affiliation{Department of Physics and Astronomy, Rutgers University, Piscataway, NJ 08854, USA}

\begin{abstract}
Superconducting cavities with high quality factors, coupled to a fixed-frequency transmon, provide a state-of-the-art platform for quantum information storage and manipulation.
The commonly used selective number-dependent arbitrary phase (SNAP) gate faces significant challenges in ultra-high-coherence cavities, where the weak dispersive shifts necessary for preserving high coherence typically result in prolonged gate times. 
Here, we propose a protocol to achieve high-fidelity SNAP gates that are orders of magnitude faster than the standard implementation, surpassing the speed limit set by the bare dispersive shift. 
We achieve this enhancement by dynamically amplifying the dispersive coupling via sideband interactions, followed by quantum optimal control on the Floquet-engineered system. 
We also present a unified perturbation theory that explains both the gate acceleration and the associated benign drive-induced decoherence, corroborated by Floquet\textendash Markov simulations. 
These results pave the way for the experimental realization of high-fidelity, selective control of weakly coupled, high-coherence cavities, and expanding the scope of optimal control techniques to a broader class of Floquet quantum systems.
\end{abstract}

\maketitle

\section{Introduction}\label{sec:intro}
Circuit quantum electrodynamics (cQED)~\cite{blais_rmp}, comprising superconducting circuits coupled to microwave cavities, is one of the leading platforms for quantum information processing.  
Traditionally, quantum information is encoded in two levels of a superconducting circuit, forming a qubit, while a microwave cavity primarily serves as a readout element~\cite{DiVincenzo}.  
The development of three-dimensional (3D) cavities with millisecond coherence enabled the use of bosonic modes for quantum memory and processing—an approach further strengthened by recent advances pushing coherence times beyond tens of milliseconds~\cite{coh_1,coh_2,coh_3}.
The large Hilbert space of cavity modes offers several distinct advantages over multi-qubit architectures for quantum chemistry~\cite{PhysRevX.10.021060,MacDonell2021,PhysRevX.13.011008,Girvin2024hybrid} and quantum simulation~\cite{Rico2018, crane2024hybrid}.  
Additionally, bosonic error correction codes, where information is redundantly encoded in the cavity modes, have shown promising progress in reaching the break-even point~\cite{Ofek2016-jd,Hu2019-kg,campagne2020quantum,Gertler2021-hq,ni2023beating, Sivak2023-py}. 

To manipulate information stored in a bosonic mode, a nonlinear transmon ancilla is typically coupled to the cavity in the dispersive regime~\cite{dispersive_decoherence}, characterized by the dispersive shift $\chi$. While this coupling enables universal control, it also leads to Purcell-induced cavity decay and unwanted nonlinearities from the lossy transmon—issues that hinder the efficacy of bosonic quantum error correction~\cite{campagne2020quantum,Sivak2023-py} and become increasingly significant as cavity coherence times improve. Reducing $\chi$ can suppress these inherited effects, but the resulting weaker coupling lengthens gate times and can potentially limit gate fidelities due to ancilla decoherence. 
Consequently, several gate schemes have been developed to accelerate operations at weak coupling. The echoed conditional displacement (ECD) gate~\cite{Eickbusch2021} and the conditional-not displacement (CNOD) gate~\cite{Diringer2023} enhance the effective coupling through large cavity displacements, while sideband-based gates~\cite{sideband,fault_tolerant_snap,Chakram2020,huang2025fastsidebandcontrolweakly} achieve similar speedups via transmon displacements.
In contrast, accelerating the selective number-dependent arbitrary phase (SNAP) gate~\cite{snap_1,snap_2}\textemdash one of the most widely used cavity control protocols\textemdash remains challenging.
Nevertheless, SNAP gates play a critical role in enabling more efficient algorithmic decompositions for key applications~\cite{Fosel2020,kurkccuoglu2024qudit}, and thus remain an essential primitive for bosonic quantum computation, motivating efforts to achieve faster SNAP gates. 

In prior work, faster SNAP gates have typically been realized using optimal control techniques that refine the pulse shape to minimize leakage and infidelity~\cite{Kudra2022}. More recently, it was shown that coherent errors associated with short-duration SNAP pulses can be completely suppressed, provided the pulse duration exceeds a critical threshold~\cite{fast_snap}. Despite these advances, SNAP gate times remain constrained by the fundamental limit set by the dispersive shift. This constraint arises because the SNAP drive must possess a narrow spectral bandwidth, precisely centered on the target transition, to enable photon-number–selective transmon control.

In this work, we propose a fast and high-fidelity SNAP gate that surpasses this fundamental speed limit.  
First, we engineer sideband interactions to dynamically enhance the effective dispersive shift, adapting a mechanism previously used for its suppression~\cite{fault_tolerant_snap,fault_tolerant_2,Reinhold2020-ho}, and achieve an order-of-magnitude increase.
We then develop a protocol to implement the SNAP gate in the Floquet basis defined by the sideband drive, whose effects can be understood analytically.  
We refer to this as the Floquet SNAP gate, which yields a tenfold acceleration in gate speed over the standard SNAP gate.
Furthermore, we extend optimal control techniques~\cite{Heeres2017,khaneja2005optimal} to the Floquet-engineered SNAP gate, achieving an additional order of magnitude speedup while maintaining high fidelity. 
Finally, we identify the dominant drive-induced decoherence mechanisms and verify through open-system simulations that they do not negate the advantages of the accelerated gate.  

This paper is organized as follows: 
Section~\ref{sec:snap_sideband} reviews the basics of the SNAP gate and introduces sideband engineering as a method to enhance the dispersive shift.
Section~\ref{sec:floquet_snap} develops a theoretical framework for implementing the SNAP gate in the Floquet basis and presents numerical simulations demonstrating a high-fidelity operation with reduced gate time.
Section~\ref{sec:qoc_snap} extends quantum optimal control techniques to the Floquet SNAP gate, achieving further acceleration and fidelity improvement.  
The influence of decoherence on the Floquet SNAP gate is evaluated through open-system simulations in Sec.~\ref{sec:noise}, where a comparative study highlights the overall fidelity improvements enabled by this approach.
Finally, Sec.~\ref{sec:con_discussion} summarizes the work and explores potential applications, with additional details and derivations provided in the Appendices.

\section{SNAP gate and sideband engineering}\label{sec:snap_sideband}

\begin{figure*}
    \centering
    \includegraphics[width=\textwidth]{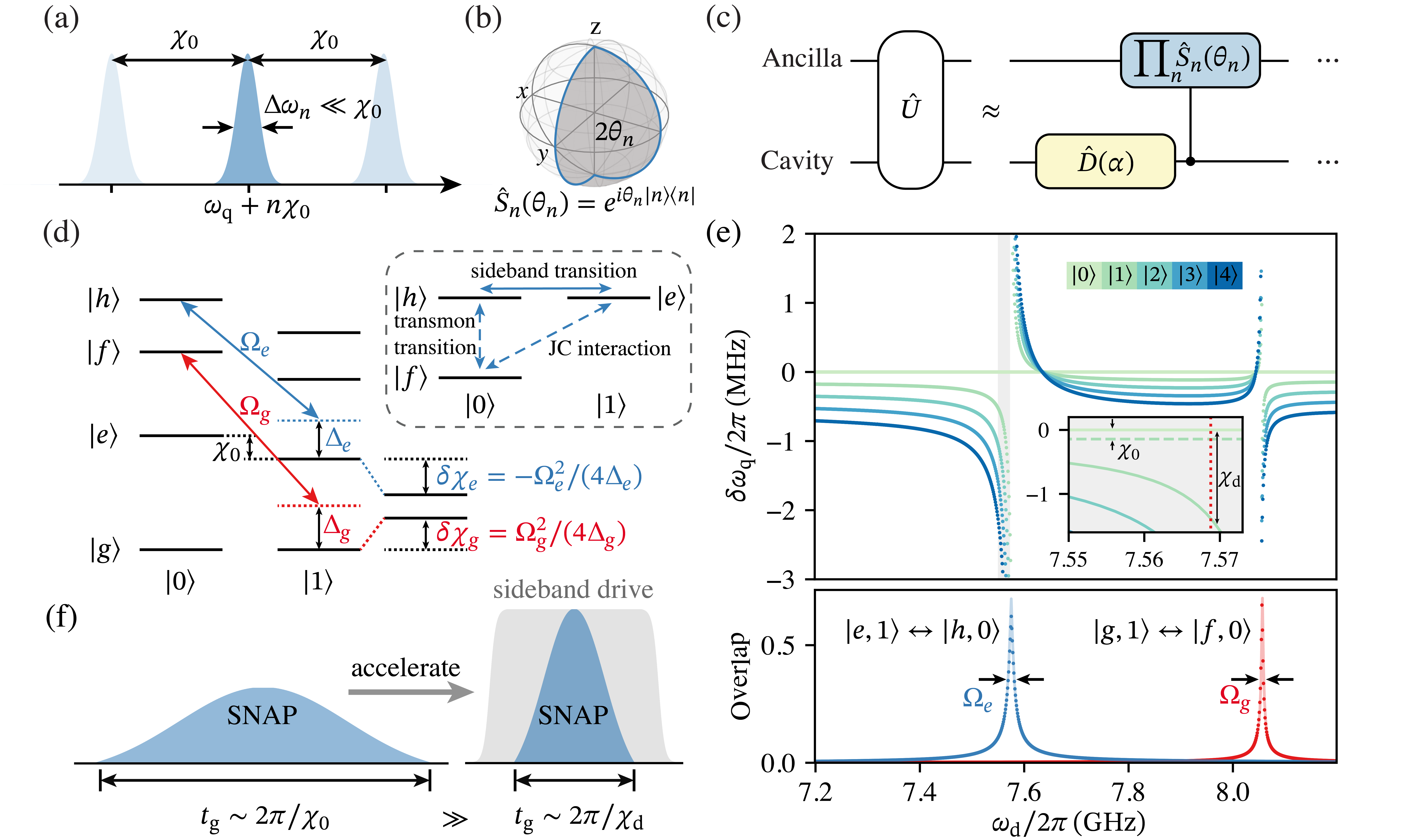}
    \caption{
    Accelerating SNAP gates by boosting the dispersive shift through sideband engineering.
    (a) Spectrum of a standard ancilla drive in a SNAP gate. Each peak is centered at $\omega_\text{q} + n\chi$ with linewidth constrained by $\Delta\omega_n \ll |\chi|$.
    (b) Bloch-sphere trajectory of the ancilla under a drive conditioned on the cavity state $|n\rangle$. The enclosed geometric phase $2\theta_n$ imparts a relative phase $\theta_n$ onto the Fock state $|n\rangle$.
    (c) Any arbitrary unitary $\hat{U}$ for a single cavity mode can be decomposed into a sequence of displacement operations $\hat{D}(\alpha)$ interleaved with SNAP gates $\prod_n \hat{S}_n(\theta_n)$.
    (d) Energy-level diagram of the system. The blue and red arrows denote the sideband transitions $|g,n+1\rangle \leftrightarrow |f,n\rangle$ and $|e,n+1\rangle \leftrightarrow |h,n\rangle$, with amplitudes $\Omega_{g,e}$ and detunings $\Delta_{g,e}$. The resulting dispersive shifts induced by the sideband interactions are labeled $\delta\chi_{g,e}$. 
    The inset (in the rotating frame of the drive frequency) illustrates that the sideband transition arises from a Raman-like process mediated by the Jaynes\textendash Cummings interaction and a single-photon transmon transition.
    (e) Ancilla frequency shift $\delta\omega_\text{q}$ as a function of drive frequency for different cavity photon numbers. The inset zooms in near the $|e,n+1\rangle \leftrightarrow |h,n\rangle$ sideband resonance, showing both the bare dispersive shift $\chi_0$ and the driven dispersive shift $\chi_\text{d}$ at the selected drive frequency (red dashed line). Bottom panel shows the overlap between Floquet and static eigenstates. Blue and red dots correspond to $_\text{S}\langle e,1|h,0\rangle_\text{F}$ and $_\text{S}\langle g,1|f,0\rangle_\text{F}$, respectively, with fits to Eq.~\eqref{eq:fit} shown as transparent curves. The linewidth of each peak reflects the sideband coupling strength.
    (f) Schematic illustration of SNAP gate acceleration. The SNAP pulse duration is shortened by applying a background sideband drive that enhances the dispersive shift.
    Parameters: $\omega_\text{q}/2\pi = 6.4$ GHz, $\omega_\text{c}/2\pi = 4.5$ GHz, $\alpha/2\pi = -230$ MHz, $\chi_0/2\pi = 0.14$ MHz, $\epsilon/2\pi = 0.8$ GHz.}
\label{fig:FIG_OVERVIEW}  
\end{figure*}

We start by providing a brief review of the SNAP gate, emphasizing constraints on the gate time imposed by the dispersive shift. 
We then revisit the use of sideband driving as a technique to tune the dispersive shift, potentially enabling faster gate operations.

\subsection{Selective number-dependent arbitrary phase (SNAP) gate}

The selective number-dependent arbitrary phase (SNAP) gate is one of the most widely used protocols for controlling cavity modes~\cite{snap_1,snap_2}.
In combination with cavity displacements $\hat{D}(\alpha)$, SNAP gates enable universal control of a single cavity mode. This protocol operates in the dispersive regime of a cavity–ancilla system, where the two modes are far-detuned relative to their coupling strength, and is described by the Hamiltonian
\begin{equation}
    \hat{H}_\text{disp} = \omega_\text{c}\hat{a}^\dag\hat{a} + \omega_\text{q}\hat{q}^\dag\hat{q} + \frac{\alpha}{2}\hat{q}^{\dag 2}\hat{q}^2
    + \chi \hat{a}^\dag\hat{a}\hat{q}^\dag\hat{q},
\end{equation}  
where \( \hat{a} \) and \( \hat{q} \) are the ladder operators of the cavity and ancilla, with frequencies \( \omega_\text{c} \) and \( \omega_\text{q} \), respectively. The anharmonicity \( \alpha \) allows the ancilla to be treated as a two-level system under sufficiently weak microwave drives.  
The last term in the above Hamiltonian represents a dispersive interaction, where the ancilla frequency shifts depending on the number of photons in the cavity (or vice versa), enabling selective control over individual Fock states.  
A SNAP gate is typically implemented by driving the ancilla with a set of Gaussian pulses, applied either sequentially or simultaneously as a frequency comb, where the \( n \)th pulse is centered at \( \omega_\text{q} + n\chi \) [see Fig.~\ref{fig:FIG_OVERVIEW}(a)].  
When the linewidth of each pulse \( \Delta \omega_n \) is much smaller than the dispersive shift \( \chi \), the \( n \)th pulse selectively excites the ancilla conditioned on the cavity being in the Fock state \( |n\rangle \).  
Each such pulse induces two consecutive $\pi$ rotations with rotation axes differing by $\theta_n$ on the equator.
The resulting closed-loop trajectory of the ancilla on the Bloch sphere, encloses a geometric phase \( 2\theta_n \), which imparts a relative phase shift to the corresponding Fock state:  $\hat{S}_n(\theta_n) = e^{i\theta_n |n\rangle\langle n|}$ [see Fig.~\ref{fig:FIG_OVERVIEW}(b)]. 
Driving the ancilla at frequencies corresponding to different photon numbers enables arbitrary phase shifts across selected Fock states, thereby realizing the SNAP gate \( \prod_n \hat{S}_n(\theta_n) \).  
To realize universal control over a single cavity mode, any arbitrary unitary operations can be decomposed into a sequence of SNAP and displacement gates, as shown in Fig.~\ref{fig:FIG_OVERVIEW}(c).

Compared to other cavity control protocols that are unselective, such as the echoed-conditional displacement (ECD) gate~\cite{Eickbusch2021,you2024} and the conditional-not displacement (CNOD) gate~\cite{Diringer2023}, the SNAP gate offers greater flexibility, enabling independent tuning of phases associated with individual Fock states.
This additional control may allow for more efficient gate decompositions in certain quantum algorithms~\cite{kurkccuoglu2024qudit}.
However, achieving high-fidelity operations with negligible coherent error often requires gate durations substantially longer than the associated fundamental timescale \( 2\pi/\chi \).  
Although optimal control techniques can reduce this duration, approaching the \( 2\pi/\chi \) limit remains challenging.  
Given the limited coherence time of the ancilla, the relatively long duration of the SNAP gate remains a limiting factor for cavity control fidelity, especially in ultra-high coherence cavity systems with necessarily small dispersive shifts.

\subsection{Tunable dispersive shift from sideband engineering}

One approach to accelerating the SNAP gate beyond its fundamental speed limit is to temporarily enhance the dispersive shift during gate operation.  
Physically, the dispersive shift arises from the Jaynes\textendash Cummings interaction, \( g(\hat{a}^\dag\hat{q}+\hat{a}\hat{q}^\dag) \), which facilitates photon exchange between the cavity and ancilla modes, and results in a photon-number-dependent frequency shift of the ancilla when the two are far detuned. Similarly, sideband interactions~\cite{sideband} lead to analogous photon-number-dependent shifts when activated off-resonantly~\cite{Reinhold2020-ho}.
Within the dispersive regime, sideband interactions can be induced by driving the ancilla at specific frequencies.  
Qualitatively, these interactions can be interpreted as a second-order cavity-assisted Raman process~\cite{sideband}. For example, the $|e1\rangle \leftrightarrow |h0\rangle$ sideband is mediated by the Jaynes\textendash Cummings interaction $|e1\rangle \leftrightarrow |f0\rangle$ and the single-photon ancilla transition $|f0\rangle \leftrightarrow |h0\rangle$ [see the inset in Fig.~\ref{fig:FIG_OVERVIEW}(d)].

To quantitatively describe this effect, we perform a dispersive transformation on the ancilla drive term, \( \epsilon \cos(\omega_\text{d}t) (\hat{q}^\dag + \hat{q}) \), leading to the following effective interaction (see Appendix~\ref{APP:DERIVATION_DISPERSIVE} for a detailed derivation):  
\begin{equation}
   \hat{H}_\text{sb} = \sum_{j=g,e} \frac{\Omega_j}{2}(\hat{a}^\dag |j\rangle \langle j+2|e^{-i\Delta_j t} + \text{h.c.}),
\end{equation}  
where \( \Omega_j \) and \( \Delta_j \) represent the coupling strength and detuning of the \( j \)th sideband transition.  
In the remainder of this work, we focus on two specific sideband transitions:  \( |g,n+1\rangle \leftrightarrow |f,n\rangle \) and \( |e,n+1\rangle \leftrightarrow |h,n\rangle \).  
Defining the detunings \( \Delta = \omega_\text{q} - \omega_\text{c} \) and \( \Delta_\text{q} = \omega_\text{q} - \omega_\text{d} \), the corresponding sideband rates are given by  $\Omega_g = -\frac{\sqrt{2} \epsilon g \alpha}{\Delta (\Delta+\alpha)}$  and $\Omega_e = -\frac{\sqrt{6} \epsilon g \alpha}{(\Delta+\alpha)(\Delta+2\alpha)}$,
with detunings  $\Delta_g = \Delta+\Delta_\text{q}+\alpha$ and $ \Delta_e = \Delta+\Delta_\text{q}+2\alpha$.

In the regime \( \Delta_j \gg \Omega_j \), these sideband interactions lead to energy shifts, analogous to the static dispersive shift in the absence of a drive.
There, we apply a Schrieffer–Wolff transformation (SWT) to diagonalize the sideband Hamiltonian \( \hat{H}_\text{sb} \), with the generators $\hat{G}_j  =  \frac{\Omega_j}{2\Delta_j}e^{i\Delta_j t}\hat{a}|j+2\rangle \langle j| - \text{h.c.}$
This leads to a driven Hamiltonian analogous to the static dispersive Hamiltonian~\cite{fault_tolerant_snap}:  
\begin{equation*}
    \hat{H}_\text{sb}' = \sum_{j=g,e} \bigg(-\frac{\Omega_j^2}{4\Delta_j} \hat{a}^\dag \hat{a} |j\rangle \langle j| + \frac{\Omega_j^2}{4\Delta_j} \hat{a}^\dag \hat{a} |j+2\rangle \langle j+2|\bigg).
\end{equation*}  
Depending on the drive frequency, one of the sideband interactions dominates, leading to a modified dispersive shift:  
\begin{equation}\label{eq:disp_2}
    \chi_\text{d}^{g,n+1\leftrightarrow f,n} = \chi_0 +  \frac{\Omega_g^2}{4\Delta_g}, \quad \chi_\text{d}^{e,n+1\leftrightarrow h,n} = \chi_0 -
    \frac{\Omega_e^2}{4\Delta_e} .
\end{equation}  
A schematic energy-level diagram for the sideband interactions and dispersive shift corrections is shown in Fig.~\ref{fig:FIG_OVERVIEW}(d).

We verify these analytical results with numerical calculations of the Floquet quasienergy as a function of the drive frequency \( \omega_\text{d} \), as shown in Fig.~\ref{fig:FIG_OVERVIEW}(e). The colored curves, transitioning from light to dark, indicate the shift in the ancilla transition frequency when the cavity is in different Fock states $|n\rangle$, with the zero-photon case serving as a baseline.  
In particular, the energy difference between the cavity in \( |0\rangle \) and \( |1\rangle \) defines the driven dispersive shift \( \chi_\text{d} \). Notably, \( \chi_\text{d} \) varies as a function of \( \omega_\text{d} \) and exhibits divergences at two specific frequencies, corresponding to the \( |g,n+1\rangle \leftrightarrow |f,n\rangle \) and \( |e,n+1\rangle \leftrightarrow |h,n\rangle \) resonant sideband transitions.  
The inset provides a zoomed-in view near the \( |e,n+1\rangle \leftrightarrow |h,n\rangle \) resonance. The red dotted line marks the drive frequency used in the rest of this paper, highlighting that the driven dispersive shift \( |\chi_\text{d}|/2\pi = 1.4 \) MHz is approximately ten times larger than the undriven dispersive shift \( |\chi_0|/2\pi = 0.14 \) MHz.

To numerically extract the sideband rate and detuning, a direct fit to the dispersive shifts in Eq.~\eqref{eq:disp_2} is inaccurate, as the equation is only valid in the sideband-dispersive regime (\( \Delta_i \gg \Omega_i \)).  
Instead, we determine these parameters by fitting the overlap between the Floquet and static eigenstates (denoted by subscripts F and S), shown as dots in the lower panel of Fig.~\ref{fig:FIG_OVERVIEW}(e).  
The blue and red curves correspond to the overlaps \( _\text{S}\langle e, 1| h,0\rangle _\text{F} \) and \(_\text{S} \langle g, 1| f,0\rangle _\text{F} \), respectively.  
We then fit to the following expression~\cite{schuster2007circuit}, which is valid beyond the dispersive regime: 
\begin{equation}\label{eq:fit}
     _\text{S}\langle j, 1| j+2,0\rangle _\text{F}  = \sin\left[\frac{1}{2}\arctan\left(\frac{\Omega_j}{\Delta_j}\right)\right].
\end{equation}  
The fitted results are shown as solid lines in the lower panel of Fig.~\ref{fig:FIG_OVERVIEW}(e).  
For instance, the red dotted line in the upper panel corresponds to a detuning of \( \Delta_e/2\pi = 6.8 \) MHz and a sideband rate of \( \Omega_e/2\pi = 5.3 \) MHz.  
Applying Eq.~\eqref{eq:disp_2} yields a predicted dispersive shift of 1.2 MHz, slightly lower than the 1.4 MHz obtained from exact diagonalization.  
This discrepancy arises from higher-order perturbative effects, which are captured by the Floquet analysis.

\section{The Floquet SNAP gate}\label{sec:floquet_snap}

\begin{figure*}[t]
\centering
\includegraphics[width=\textwidth]{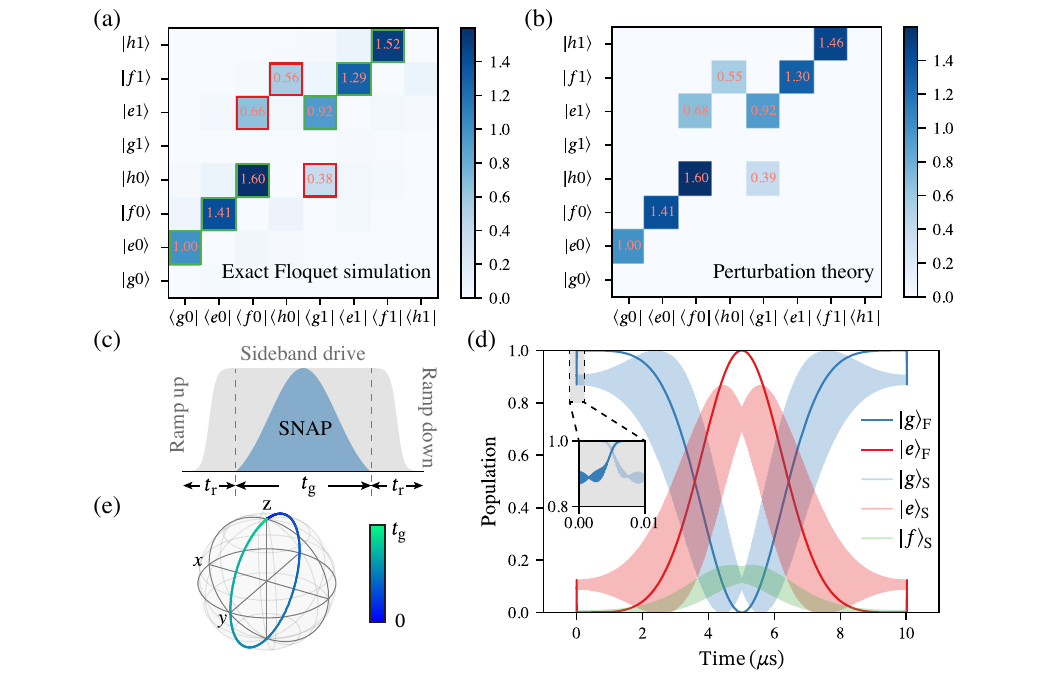}
\caption{
Implementing the SNAP gate in the Floquet basis.
(a) Floquet matrix elements of the ancilla ladder operator $\hat{q}$, maximized over the Floquet band index $k$. Red boxes highlight elements that are not present in the static basis.
(b) The same quantities as in (a), obtained through a perturbative treatment of the sideband interactions.
(c) Pulse sequence for Floquet SNAP (not to scale). The gray background represents the sideband drive, consisting of a ramp-up, flat-top, and ramp-down segment. The Gaussian SNAP pulse is shown in blue.
(d) Population dynamics of various ancilla states, where solid and transparent curves correspond to eigenstates in the Floquet basis and dressed basis, respectively. The inset provides a zoomed-in view of the ramp-up region, illustrating the smooth transition from the static to Floquet eigenstate.
(e) Trajectory of the ancilla state evolution on the Bloch sphere within the Floquet basis.
Parameters: $\omega_\text{q}/2\pi=6.4$ GHz, $\omega_\text{c}/2\pi=4.5$ GHz, $\alpha/2\pi=-230$ MHz, $\chi_0/2\pi=0.14$ MHz, $\epsilon/2\pi = 0.8$ GHz, $\omega_\text{d}/2\pi = 7.56$ GHz, $t_\text{r} = 10$ ns, $t_\text{g} = 10$ µs. }
\label{fig:FIG_FLOQUET_SNAP}
\end{figure*}

In the previous section, we showed that sideband engineering enables a significant enhancement of the dispersive shift. A natural subsequent question is whether this enhancement can be harnessed to accelerate the implementation of SNAP gates, as schematically illustrated in Fig.~\ref{fig:FIG_OVERVIEW}(f). Indeed, the answer is affirmative. In this section, we outline a protocol for realizing fast and high-fidelity SNAP gates by leveraging the dynamically enhanced dispersive shift. 

Quantitatively, the question raised above pertains to the broader context of gate operations in Floquet systems, a topic that has attracted significant attention in recent years. Notable examples include engineering dynamical flux sweet spots in fluxonium qubits~\cite{floquet_thy,floquet_exp,floquet_fm}, longitudinal readout schemes for transmon qubits~\cite{floquet_readout_1,floquet_readout_2}, and programming tunable Heisenberg spin interactions for quantum simulations~\cite{Nguyen2024}.  
A Floquet system is defined by a time-periodic Hamiltonian \( \hat{H}_\text{F}(t+T) = \hat{H}_\text{F}(t) \), with period \( T = 2\pi/\omega_\text{d} \).  Solutions to the corresponding Schrödinger equation \( i\hbar \partial_t \Psi_j(t) =  \hat{H}_\text{F}(t) \Psi_j(t) \) take the general form \( \Psi_j(t) = e^{-i\varepsilon_j t} \psi_j(t) \), with \( \Psi_j(t) \) denoting Floquet states, \( \psi_j(t) = \psi_j(t+T) \) the Floquet modes, and \( \varepsilon_j \) the Floquet quasienergies.
Unlike in static systems, the Floquet quasienergies and their modes are only uniquely defined within a given Brillouin zone, e.g., \( \varepsilon \in [0, \omega_\text{d}] \). 
Consequently, a Floquet mode \( \psi_j(t) \) with quasienergy \( \varepsilon_i \) is physically equivalent to the mode \( \psi_j(t) e^{ik\omega_\text{d} t} \) with quasienergy \( \varepsilon_j + k\omega_\text{d} \) for any \( k \in \mathbb{Z} \), since  
\( \Psi_j(t) = e^{-i\varepsilon_j t} \psi_j(t) = e^{-i(\varepsilon_j+k\omega_\text{d}) t} \psi_j(t)e^{ik\omega_\text{d} t} \).  
While this Brillouin-zone structure is a mathematical redundancy, it plays a crucial role in interpreting physical phenomena such as multiphoton resonances which involve transitions between Floquet modes in different Brillouin zones.  
A more comprehensive review of Floquet physics can be found, for example, in the work of Grifoni and Hänggi~\cite{hanggi1991}.

To engineer a unitary operation in a Floquet system, such as a SNAP gate, it is essential to determine the relevant transition frequencies and matrix elements.
In Fig.~\ref{fig:FIG_FLOQUET_SNAP}(a), we numerically compute the Floquet matrix elements of the ancilla annihilation operator \( \hat{q} \) (related to the drive operator \( \hat{q}+\hat{q}^\dag \)), maximized over the Brillouin-zone index \( k \),  
\begin{equation}
    M_{ij} = \max_k \frac{1}{T}\int_0^T \mathrm{d}t\, e^{-ik\omega_\text{d}t}\langle i(t)|\hat{q} | j(t)\rangle,
\end{equation}
where each index $i, j$ labels a composite state $|m, n\rangle$ of the ancilla and cavity, with the ancilla state listed first and the cavity state second.
The magnitudes of the matrix elements exceeding 0.1 are explicitly displayed.  
Two key differences arise compared to the static case.   
First, single-photon transitions in the ancilla, marked by green boxes in Fig.~\ref{fig:FIG_FLOQUET_SNAP}(a), now depend on the number of photons in the cavity.  
This dependence arises from a second-order perturbative correction to the ancilla ladder operator due to the sideband interaction (see Appendix~\ref{APP:DERIVATION_SIDEBAND} for details),  
\begin{equation}\label{eq:side_flo_1}
    \hat{q}(t) + \frac{1}{2}[\hat{G},[\hat{G},\hat{q}(t)]] = \hat{q}(t)
    -\frac{1}{8} \left( \frac{\Omega}{\Delta} \right)^2 \hat{a}^\dag \hat{a}
    \hat{q}(t).
\end{equation}
We drop the subscript on the sideband rate and detuning, as only the \( |e,n+1\rangle \leftrightarrow |h,n\rangle \) transition is considered in the remainder of this paper.  
Second, additional transitions emerge wherein the ancilla and cavity exchange photons, highlighted by the red boxes in Fig.~\ref{fig:FIG_FLOQUET_SNAP}(a).  
These transitions also arise from sideband interactions and are captured by first-order perturbative correction to the ancilla ladder operator,
\begin{equation}\label{eq:side_flo_2}
    \begin{aligned}
            \hat{q}(t) + [\hat{G},  \hat{q}(t)]  =& \hat{q}(t)  - \frac{\Omega_e}{2\Delta_e}\hat{a}^\dag|g\rangle\langle h|e^{-i(\Delta_e+\omega_\text{q})t} \\
   & + \frac{\sqrt{3}\Omega_e}{2\Delta_e}\hat{a}|f\rangle\langle e|e^{i(\Delta_e-\omega_\text{q}-2\alpha)t} \\ 
   & - \frac{\sqrt{2}\Omega_e}{2\Delta_e}\hat{a}|h\rangle\langle f|e^{-i(\Delta_e-\omega_\text{q}-\alpha)t}. 
    \end{aligned}
\end{equation}
Analytical predictions from Eqs.~\eqref{eq:side_flo_1} and~\eqref{eq:side_flo_2} are plotted in Fig.~\ref{fig:FIG_FLOQUET_SNAP}(b), showing excellent agreement with numerical results from Fig.~\ref{fig:FIG_FLOQUET_SNAP}(a).  
In addition to the matrix elements, transition frequencies for processes \( |i\rangle \to |j\rangle \) in Fig.~\ref{fig:FIG_FLOQUET_SNAP}(a) are obtained from quasienergy differences and the drive frequency multiplied by the Floquet index \( k_\text{max} \),  i.e., $\omega_{ij} = \varepsilon_j - \varepsilon_i -k_\text{max}\omega_\text{d}$.
The presence of Floquet multipliers is consistent with the perturbative results in Eq.~\eqref{eq:side_flo_2}.  
For example, the transition frequency for \( \hat{a}^\dag | g\rangle \langle h| \) is \( \Delta_e + \omega_\text{q} = (3\omega_\text{q} + 3\alpha -\omega_\text{c}) - \omega_\text{d} \), where the terms in parentheses denote the transition frequency in the static case, while the last term arises due to $k_\text{max}=1$.

Having identified the transition frequencies and matrix elements, we now proceed to implement a SNAP gate in the Floquet basis.  
As a concrete example, we consider the SNAP gate \( e^{i\pi|0\rangle \langle 0|} \), which imparts a \( \pi \) phase shift exclusively to the \( |0\rangle \) cavity state, leaving all other states unaffected.  
Figure~\ref{fig:FIG_FLOQUET_SNAP}(c) illustrates the applied pulse sequence. 
The sideband drive, indicated by the gray background, has amplitude \( \epsilon \) and frequency \( \omega_\text{d} \).  
To ensure an adiabatic mapping between states in the Floquet basis and the lab frame, the drive is smoothly ramped up and down over a duration \( t_\text{r} = 10 \) ns (see Appendix~\ref{APP:RAMP} for the choice of ramp time). 
During the flat-top portion of the sideband drive pulse, we simultaneously apply an additional Gaussian pulse with duration $4\sigma$ (shown in blue) to induce the ancilla rotation required for the SNAP gate. 
The carrier frequency of the Gaussian is set to the sideband-dressed transition frequency \( \omega_{e0, g0} \), and the amplitude is determined by the gate time and the matrix element \( M_{e0, g0} \), ensuring the integrated pulse area is \( 2\pi \).  
We choose a SNAP gate time of \( t_\text{g} = 10 \) µs, which is approximately 1.5 times the fundamental timescale set by the undriven dispersive shift $\chi_0$, and significantly shorter than the typical SNAP duration of nearly ten times the same scale.

Figure~\ref{fig:FIG_FLOQUET_SNAP}(d) shows the evolution of the ancilla population when the system is initialized in the static (undriven) eigenstate \( |g,0\rangle_\text{S} \).  
Solid and transparent curves represent the population in the Floquet and the static bases, respectively.  
The inset zooms in on the ramp-up period, illustrating the smooth transition of the ancilla state from the static eigenstate \( |g\rangle_\text{S} \) to the Floquet eigenstate \( |g\rangle_\text{F} \).  
Subsequently, the ancilla population evolves as expected within the Floquet basis, transitioning from \( |g\rangle_\text{F} \) to \( |e\rangle_\text{F} \) and then back to \( |g\rangle_\text{F} \), with its Bloch sphere trajectory shown in Fig.~\ref{fig:FIG_FLOQUET_SNAP}(e).  
At the end of the pulse sequence, the ramp-down pulse maps the Floquet eigenstate \( |g\rangle_\text{F} \) back to the static eigenstate \( |g\rangle_\text{S} \). For reference, the evolution in the static basis is also shown, revealing population in the lowest three states oscillating at a carrier frequency of \( 2\omega_\text{d} \).

To systematically evaluate the performance of the Floquet SNAP gate, we compute its gate fidelity.  
We truncate the cavity Hilbert space of the system to the lowest \( N_\text{c} \) photon states, with \( N_\text{c} \) determined by the specific application.  
Here, we take \( N_\text{c} = 5 \).  
We first construct the reference propagator \( \hat{U}_\text{ref} \), defined in the presence of only the sideband drive.  
Since the ancilla always starts in the \( |g\rangle \) state for a SNAP gate, we solve the Schrödinger equation for each eigenstate \( |g,n\rangle \) within the truncated Hilbert space to obtain their final states \( |g,n; t=t_\text{g}\rangle \).  
The matrix elements of \( \hat{U}_\text{ref} \) in this eigenbasis are thus given by 
\begin{equation}
    \hat{U}_\text{ref}^{(i,j)} = \langle g,i| g,j; t=t_\text{g}\rangle.
\end{equation}
Similarly, we construct the propagator \( \hat{U} \) when both the sideband and SNAP drives are applied.  
The propagator in the interaction picture, defined relative to \( \hat{U}_\text{ref} \), is given by  
$\hat{U}_\text{int} = \hat{U}_\text{ref}^\dag \hat{U}$.
Using the target propagator \( \hat{U}_\text{target} = \text{Diag}(e^{i\pi},1,1,1,1,1) \), the gate fidelity is then evaluated as  
$\text{Tr}(\hat{U}_\text{target}^\dag \hat{U}_\text{int})^2/N_\text{c}^2$.
The calculated fidelity for the 10-µs Floquet SNAP gate is $99.8\%$. For comparison, a standard SNAP gate under the same duration and system parameters yields a simulated fidelity of only $84.3\%$. This reduction in infidelity by two orders of magnitude clearly underscore the advantage of the Floquet SNAP, highlighting its capability to achieve fast and high-fidelity cavity control that significantly surpasses the standard approach.

\section{Optimal control of Floquet SNAP gate}\label{sec:qoc_snap}

\begin{figure*}[t]
\centering
\includegraphics[width=\textwidth]{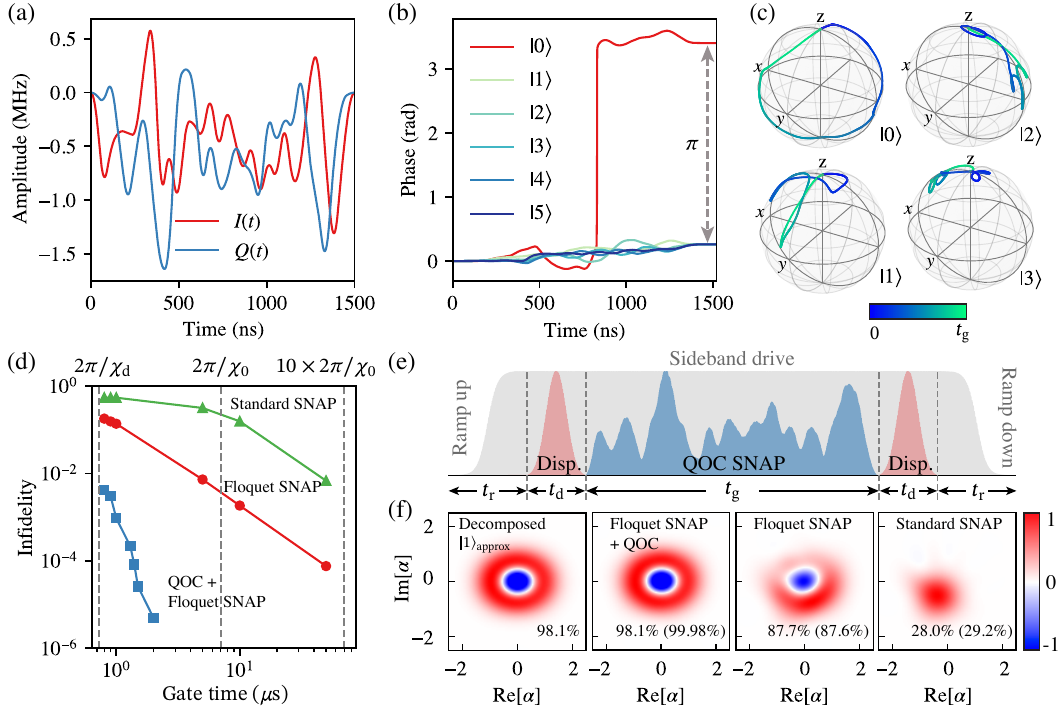}
\caption{
Optimal control pulses for high-fidelity SNAP gates in the Floquet basis. 
(a) \( I \) and \( Q \) quadratures of an optimized pulse implementing the \(\exp(i\pi|0\rangle\langle0|)\) SNAP gate in 1500 ns. 
(b) Phase accumulation dynamics for various Fock states, demonstrating the desired \( \pi \)-phase difference between the target \( |0\rangle \) state and all other states.  
(c) Ancilla trajectories on the Bloch sphere for selected cavity Fock states, illustrating the photon-number-dependent evolution induced by the optimized pulse.
(d) SNAP gate infidelity versus gate duration. Results for standard SNAP (green), Floquet SNAP (red), and QOC Floquet SNAP (blue) are shown for comparison. Dashed vertical lines mark characteristic timescales associated with the undriven dispersive shift and driven dispersive shift.
(e) Pulse sequence for Fock state \( |1\rangle \) preparation (not to scale). The gray background represents the sideband drive, consisting of ramp-up, flat-top, and ramp-down segments. The QOC SNAP pulse is shown in blue, sandwiched between two cavity displacement pulses depicted in red. 
(f) Wigner function of the Fock \( |1\rangle \) state, prepared using 1500-ns pulses obtained from different methods. As a reference, we also show the state $|1\rangle_\text{approx}$, constructed from a SNAP-gate decomposition and free from coherent errors. Fidelities with respect to the Fock \( |1\rangle \) state, along with the $|1\rangle_\text{approx}$ state (in parentheses), are provided.  
Parameters: $\omega_\text{q}/2\pi=6.4$ GHz, $\omega_\text{c}/2\pi=4.5$ GHz, $\alpha/2\pi=-230$ MHz, $\chi_0/2\pi=0.14$ MHz, $\epsilon/2\pi = 0.8$ GHz, $\omega_\text{d}/2\pi = 7.56$ GHz, $t_\text{r} = 10$ ns, $t_\text{d} = 72$ ns, $t_\text{g}=1500$ ns.  
}
\label{fig:FIG_QOC_FLOQUET_SNAP}
\end{figure*}  

Optimal control techniques are widely used in standard SNAP gates to achieve faster gate operation~\cite{khaneja2005optimal,Kudra2022,Heeres2017}. In this section, we investigate the feasibility of extending optimal control to the Floquet SNAP gate to further accelerate the gate speed and improve fidelity.  

As discussed in Sec.~\ref{sec:snap_sideband}, the primary limitation on SNAP gate speed originates from its photon-number-selective nature, where each pulse contains a frequency component that targets a specific cavity Fock state.  
Generally, reducing the gate time broadens the pulse spectral bandwidth, inadvertently exciting nearby transitions involving unintended cavity states. 
This constraint can be partially mitigated through optimal control techniques, such as \texttt{GRAPE}~\cite{khaneja2005optimal}.  
By explicitly accounting for the influence of the applied pulse on unwanted cavity states, these techniques allow shaping of the pulse to suppress these undesired effects.  
Rather than eliminating these spurious transitions, the optimized pulses cancel their net effect over time (up to a global phase), ensuring that the only impact is the intended phase accumulation on the target Fock states.  

Applying optimal control to Floquet SNAP gates introduces additional complexity due to the external drive simultaneously exciting the ancilla and inducing transitions between the ancilla and cavity, as described by Eq.~\eqref{eq:side_flo_2} and illustrated in Fig.~\ref{fig:FIG_FLOQUET_SNAP}(a).
To account for these effects, we construct the optimal control Hamiltonian in the interaction picture,  
\begin{equation}
    \hat{H}_\text{qoc} = A(t)  \Big( \sum_{i,j} | i \rangle \langle j |  M_{ij}  e^{-i\omega_{ij} t} + \text{h.c.} \Big),
\end{equation}  
where the summation is over all transitions within the truncated subspace, with matrix elements \( M_{ij} \) and transition frequencies \( \omega_{ij} \).  
To implement the same SNAP gate \( e^{i\pi|0\rangle \langle 0|} \) considered previously, we employ a drive represented as
\begin{equation}
    A(t) = I(t)\sin(\omega_{e0,g0} t) + Q(t)\cos(\omega_{e0,g0} t).
\end{equation}    
The pulse profiles \( I(t) \) and \( Q(t) \) are parameterized using B-spline basis functions and optimized via automatic differentiation~\cite{auto1,auto2} implemented in \texttt{JAX}~\cite{jax2018github}. 
The optimization employs a cost function based on the gate fidelity: $\text{Tr}(\hat{P}_g\hat{U}_\text{target}^\dag \hat{U}_\text{qoc})^2/N_\text{c}^2$,
where \( \hat{U}_\text{qoc} \) is the propagator obtained through evolving the Hamiltonian \( \hat{H}_\text{qoc} \), and $\hat{P}_g$ is the projector onto the ancilla ground state.

In Fig.~\ref{fig:FIG_QOC_FLOQUET_SNAP}(a), we present an example of an optimized pulse profile for a SNAP gate duration of 1500 ns. This duration is about five times shorter than the fundamental timescale set by the undriven dispersive shift \( \chi_0 \), and only twice as long as the time limit set by the driven dispersive shift \( \chi_\text{d} \).  
To evaluate the performance of the optimized pulses, we numerically solve the Schrödinger equation using the full Hamiltonian (see details in Appendix~\ref{APP:HAM}), incorporating both the sideband and the optimal control pulses.  
Figure~\ref{fig:FIG_QOC_FLOQUET_SNAP}(b) shows the phase evolution in the interaction picture for various initial cavity Fock states.
At the end of the pulse sequence, the state \( |0\rangle \) acquires a final phase \( \phi_0 \), while all other states converge toward a common phase \( \phi_{\Bar{0}} \), satisfying the desired relation  \( \phi_0 - \phi_{\Bar{0}}  = \pi \). 
For visualization, the trajectories on the Bloch sphere for selected states are shown in Fig.~\ref{fig:FIG_QOC_FLOQUET_SNAP}(c).  

To quantify the performance of the QOC Floquet SNAP gate and compare it with both Floquet SNAP and standard SNAP gates, we compute the gate fidelity as a function of gate duration ($t_\text{g}$) for all three approaches, as presented in Fig.~\ref{fig:FIG_QOC_FLOQUET_SNAP}(d).
Relevant timescales determined by the undriven and driven dispersive shifts are indicated by vertical dashed lines for reference.  
Overall, the QOC Floquet SNAP achieves the highest fidelity within the shortest gate time, demonstrating a substantial advantage over Floquet SNAP, which itself greatly outperforms the standard SNAP.  
In particular, QOC Floquet SNAP reaches \(99\%\) fidelity within a duration shorter than \(2\pi/\chi_\text{d}\). In contrast, Floquet SNAP requires a gate duration on the order of \(2\pi/\chi_0\), approximately ten times longer than \(2\pi/\chi_\text{d}\), to reach a comparable fidelity. The standard SNAP approach is even slower, necessitating roughly ten times \(2\pi/\chi_0\) to achieve the same fidelity level.
Notably, the QOC Floquet SNAP achieves a fidelity exceeding $99.999\%$ with a gate duration three times faster than \( 2\pi/\chi_0 \), demonstrating its capability to surpass the speed limit imposed by the bare dispersive shift. 
For comparison, the ECD gate can achieve an accelerated gate time of approximately $2\pi/(\beta \chi_0)$, where the cavity displacement is bounded in the dispersive approximation by $\beta_\text{max} \approx \sqrt{\alpha/(6\chi_0)}$~\cite{Eickbusch2021}. With the parameters in Fig.~\ref{fig:FIG_QOC_FLOQUET_SNAP}, this yields a maximum speedup of about 15, comparable to that obtained with the QOC Floquet SNAP.

Up to this point, we have primarily focused on an isolated SNAP gate. However, in practical applications, SNAP gates are combined with displacement operations to enable universal control of a single cavity mode.  
As a concrete example, we demonstrate the preparation of a single-photon Fock state using an approximated displacement\textendash SNAP\textendash displacement sequence, i.e., $ |1\rangle \approx |1\rangle_\text{approx} =  \hat{D}(-0.58)e^{i\pi|0\rangle\langle0|}\hat{D}(1.14)|0\rangle$.
The gate-decomposition fidelity is \(98.1\%\), serving as our reference benchmark.
While the fidelity can be systematically improved by increasing the number of displacement–SNAP layers, we adopt this simplified sequence for illustration to reduce the computational cost associated with evolving the full Hamiltonian and Liouvillian (see Appendix~\ref{APP:HAM}).
Figure~\ref{fig:FIG_QOC_FLOQUET_SNAP}(e) shows the full pulse sequence incorporating the QOC Floquet SNAP gate (shown in blue), which is enclosed by two displacement pulses (shown in red). 
The displacement pulses have a gate duration of \(t_\text{d} = 72\,\text{ns}\), ten times faster than the timescale set by \( \chi_\text{d} \), ensuring that the displacement pulse is unselective with respect to the ancilla state.  
The cavity drive is resonant with the dressed cavity frequency, with amplitude and phase chosen according to the desired displacement magnitudes.
In Fig.~\ref{fig:FIG_QOC_FLOQUET_SNAP}(f), we plot the Wigner functions of the cavity mode after executing the full pulse sequence.  
The reference case \( |1\rangle_\text{approx} \), shown for comparison, is computed from the wavefunction obtained via gate decomposition and is free of coherent error.
The Wigner function obtained using QOC Floquet SNAP closely matches this reference, yielding a state fidelity of $99.98\%$ with respect to \( |1\rangle_\text{approx} \), which itself has a fidelity of $98.1\%$ relative to the target \( |1\rangle \) state.  
In contrast, the Wigner function obtained from Floquet SNAP exhibits noticeable distortion, with fidelity dropping below $90\%$.  
Finally, the Wigner function derived from the standard SNAP with the same gate time barely retains any quantum features, with fidelity reduced to only about $30\%$.

\section{Sideband-induced decoherence}\label{sec:noise}

In previous sections, we have shown that the sideband drive dramatically enhances the dispersive shift, enabling a significant acceleration of SNAP gates when implemented in the Floquet basis.
However, external drives may modify decoherence rates or introduce additional decoherence channels, potentially compromising gate fidelity.  
In this section, we first verify through open-system simulations that drive-induced decoherence does not diminish the advantage gained by accelerated SNAP gates. 
Subsequently, we provide a detailed analysis of the dominant noise mechanisms responsible for cavity decay and dephasing, deriving analytical predictions that show excellent agreement with numerical results from the Floquet\textendash Markov master equation.

\subsection{Impact of decoherence on Fock state preparation}

\begin{figure}[t]  
\centering  
\includegraphics[width=\columnwidth]{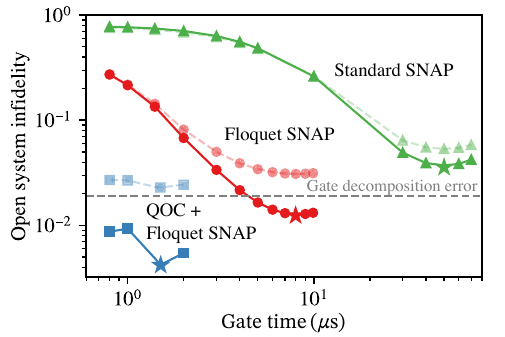}  
\caption{
Infidelity of Fock \( |1\rangle \) state preparation in the presence of decoherence.  
Infidelities obtained from standard SNAP (green), Floquet SNAP (red), and QOC Floquet SNAP (blue) gates are shown as functions of the gate duration. Solid curves represent infidelities with respect to the state from SNAP-gate decomposition $|1\rangle_\text{approx}$, while dashed transparent curves indicate infidelities relative to the target Fock state \(|1\rangle\). Stars highlight the minimum infidelity for each gate implementation. The gray dashed line indicates the intrinsic decomposition error, serving as a baseline for comparison.  
Parameters: $\omega_\text{q}/2\pi=6.4$ GHz, $\omega_\text{c}/2\pi=4.5$ GHz, $\alpha/2\pi=-230$ MHz, $\chi_0/2\pi=0.14$ MHz, $\epsilon/2\pi = 0.8$ GHz, $\omega_\text{d}/2\pi = 7.56$ GHz, $t_\text{r} = 10$ ns, $t_\text{d} = 72$ ns,  $t_\text{g}=1500$ ns, $\gamma_\text{q}=(300\,\text{µs})^{-1}$, $\gamma_{\phi,\text{q}}=(500\,\text{µs})^{-1}$, $\gamma_\text{c}=(30\,\text{ms})^{-1}$.
}  
\label{fig:FIG_FIDELITY_OPEN}  
\end{figure}  

In the following, we revisit the example of Fock state preparation studied earlier, now incorporating the effects of decoherence noise.  
Specifically, we consider an ancilla with a relaxation time of \(\gamma_\text{q}^{-1}= 300 \) µs, a pure dephasing time of \(\gamma_{\phi,\text{q}}^{-1} =500 \) µs, and a temperature of \( T = 50 \) mK.  
Additionally, the cavity is assumed to have an intrinsic decay time of $\gamma_\text{c}^{-1} =30$ ms, with negligible pure dephasing noise.  
We compute the open-system state fidelity by solving the Lindblad master equation, with results shown in Fig.~\ref{fig:FIG_FIDELITY_OPEN}.  
Here, green, red, and blue curves correspond to fidelities obtained from standard SNAP, Floquet SNAP, and QOC Floquet SNAP, respectively. The solid and dashed lines denote the fidelity with respect to the state from SNAP-gate decomposition $|1\rangle_\text{approx}$ and the target state \( |1\rangle \), respectively. For reference, the gray dashed line indicates the infidelity arising solely from gate decomposition.
As expected, including decoherence effects reduces the fidelity compared to the closed-system results in Fig.~\ref{fig:FIG_QOC_FLOQUET_SNAP}(f).
For example, with a 1500-ns pulse, the fidelity decreases from $99.98\%$ (closed system) to $99.56\%$ (open system).  
Nevertheless, the fidelities achieved by QOC Floquet SNAP remain orders of magnitude higher than those obtained with standard and Floquet SNAP approaches, outperforming them by orders of magnitude at comparable gate durations.
In addition to the state infidelity, we also evaluate the gate infidelity of the QOC Floquet SNAP, detailed in Appendix~\ref{APP:GATE_FIDELITY}. 

Analyzing fidelities as a function of gate duration reveals a non-monotonic behavior: fidelities initially improve with increased duration due to reduced coherent errors, reach a maximum (indicated by stars), and subsequently decline as decoherence dominates at longer durations. Among the three methods, QOC Floquet SNAP attains the highest maximum fidelity within the shortest gate duration.
Although Fock state preparation serves as an illustrative example, these results clearly demonstrate that the advantage of performing SNAP gates in the Floquet basis persists even in the presence of decoherence noise.  
In the remainder of this section, we identify the dominant decoherence mechanisms induced by the sideband drive and provide a systematic analysis based on both Floquet\textendash Markov simulations and analytical derivations.

\begin{figure}[t]  
\centering  
\includegraphics[width=\columnwidth]{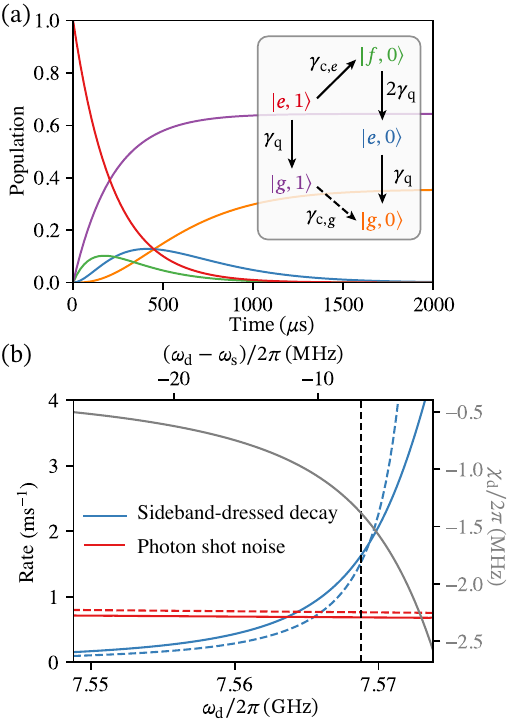}  
\caption{  
Decoherence dynamics with a sideband drive.
(a) Evolution of incoherent transitions when the system is initialized in the Floquet eigenstate \( |e,1\rangle \). The population dynamics is explained by the transition diagram (inset), which highlights key states and the incoherent transitions between them.  
(b) Dominant decoherence rates as functions of the sideband drive frequency. The blue curve represents the primary decay channel, arising from sideband dressing of the ancilla decay, while the red curve describes the leading dephasing mechanism due to photon shot noise, which results from elevated ancilla temperatures caused by dressed dephasing. Solid and dashed curves correspond to results obtained from solving the Floquet\textendash Markov equation and from the analytic expression, respectively.  
The gray curve shows the driven dispersive shift as a function of the drive frequency, included as a reference, with the drive frequency used in (a) marked by a vertical line.  
Parameters: $\omega_\text{q}/2\pi=6.4$ GHz, $\omega_\text{c}/2\pi=4.5$ GHz, $\alpha/2\pi=-230$ MHz, $\chi_0/2\pi=0.14$ MHz, $\epsilon/2\pi = 0.8$ GHz, $\omega_\text{d}/2\pi = 7.56$ GHz, $\gamma_\text{q}=(300\,\text{µs})^{-1}$, $\gamma_{\phi,\text{q}}=(500\,\text{µs})^{-1}$, $\gamma_\text{c}=(30\,\text{ms})^{-1}$.
}  
\label{fig:FIG_DECOHERENCE}  
\end{figure}

\subsection{Sideband-dressed decay}
To investigate cavity decay dynamics, we simulate the system evolution using the Floquet\textendash Markov equation (see details in Appendix~\ref{app:derivation_FM}). For simplicity, we consider only transverse coupling—i.e., coupling via $\hat{q}^\dag + \hat{q}$—between the ancilla and the bath, assuming a white-noise spectral density $J(\omega)$ with amplitude set by the intrinsic ancilla relaxation rate $\gamma_\text{q} = (300\,\text{µs})^{-1}$. Since the relevant sideband is the $|e,n+1\rangle \leftrightarrow |g,n\rangle$ transition, cavity decay is significantly enhanced when the transmon is in the excited state, with a scaling factor of $(\alpha/\Delta_e)^2$ compared to the ground-state case. Accordingly, we initialize the coupled system in the Floquet eigenstate $|e,1\rangle$.
The results, shown in Fig.~\ref{fig:FIG_DECOHERENCE}(a), exhibit nontrivial dynamics, which can be explained by the transition diagram in the inset.  
The incoherent transitions follow two primary decay pathways:  
(1) The system first undergoes a transition from \( |e,1\rangle \) to \( |f,0\rangle \) via photon exchange between the cavity and the ancilla at a rate \( \gamma_{\text{c},e} \). The ancilla then decays from \( |f,0\rangle \) to \( |e,0\rangle \) at a rate \( 2\gamma_\text{q} \), and subsequently to \( |g,0\rangle \) at a rate \( \gamma_\text{q} \);  
(2) The ancilla relaxes directly from \( |e,1\rangle \) to \( |g,1\rangle \) at a rate \( \gamma_\text{q} \), followed by cavity decay from \( |g,1\rangle \) to \( |g,0\rangle \) at a much slower rate \( \gamma_{\text{c},g} \). Since \( \gamma_{\text{c},g} \) is so small, it has a negligible effect on the dynamics within the shown time range.  
The contributions of these two decay pathways are comparable, as evidenced by the similar initial slopes of the  \( |f,0\rangle \) (magenta) and \( |g,1\rangle \) (green) population curves.  
To quantitatively extract these decay rates, we fit the population using analytical solutions to the corresponding differential equations, treating the relevant decay rates introduced above as fitting parameters.

Of particular interest is the cavity decay rate  \( \gamma_{\text{c},e} \) when the ancilla is in the excited state. Its value is extracted and plotted as a function of the sideband drive frequency in Fig.~\ref{fig:FIG_DECOHERENCE}(b) (blue solid line). The drive frequency used in Fig.~\ref{fig:FIG_DECOHERENCE}(a) is highlighted by a vertical line for reference.  
Physically, this decay channel originates from the dressing of the ancilla decay by the sideband interaction, as described by the term \( \hat{a}|f\rangle \langle e| \) in Eq.~\eqref{eq:side_flo_2}.  
Using perturbation theory, we estimate the corresponding decay rate as \( \frac{3\Omega^2}{4\Delta^2} \gamma_\text{q} \), which is plotted as the blue dashed line in Fig.~\ref{fig:FIG_DECOHERENCE}(b).  
Overall, the analytical prediction aligns well with the numerical results.  
Deviations between the analytical and numerical results become evident near the sideband resonance (as indicated by the top horizontal axis), where the perturbative approximation begins to break down. Furthermore, in deriving the analytical expression, the sideband rate \(\Omega\) is treated as a constant, though it exhibits a weak dependence on drive frequency due to cavity displacement.
As a reference, Fig.~\ref{fig:FIG_DECOHERENCE}(b) also includes the driven dispersive shift as a function of the drive frequency (gray line).  
When the drive frequency approaches the sideband resonance, a trade-off emerges: a larger dispersive shift enables faster gate operations but comes at the cost of increased cavity decay.  
For SNAP gates, which are typically limited by ancilla errors, this trade-off can be favorable, as reducing gate duration helps mitigate the impact of ancilla decoherence.
For more general applications, the drive frequency (and amplitude) can be jointly optimized to maximize overall fidelity for a given set of system parameters.

\subsection{Dressed dephasing and photon shot noise}
We examine cavity dephasing through a Ramsey-type simulation. The cavity is prepared in an equal superposition of Floquet eigenstates \( |g,0\rangle \) and \( |g,1\rangle \) and subsequently evolves under the Floquet\textendash Markov equation.  
To extract the dephasing rate, we fit the decay of the cavity coherence \(\langle 0|\, \text{Tr}_\text{q}(\rho)\, |1\rangle\), where \(\rho\) is the system density matrix. The resulting dephasing rate is plotted as the red solid line in Fig.~\ref{fig:FIG_DECOHERENCE}(b).
Unlike sideband-dressed decay, the dephasing rate exhibits weak dependence on the sideband detuning, suggesting that the dominant contribution is unrelated to sideband interactions.  
Instead, the primary source of dephasing is photon shot noise, which arises from elevated ancilla temperature due to dressed dephasing~\cite{yaxing}.  
Since dressed dephasing is only affected by the detuning between the drive frequency and the ancilla transition frequency ($\sim$GHz), it remains largely insensitive to variations in the sideband detuning ($\sim$MHz).  
To confirm this mechanism, we compute the steady-state ancilla excitation population \( n_\text{q} \) from the Floquet\textendash Markov equation and estimate the photon shot noise dephasing rate as \( n_\text{q}\gamma_\text{q} \)~\cite{shot_noise,yaxing}.  
The resulting analytical prediction is shown as the red dashed line in Fig.~\ref{fig:FIG_DECOHERENCE}(b), and shows excellent agreement with the numerically extracted dephasing rate. This confirms photon shot noise as the dominant mechanism responsible for cavity dephasing in the presence of the sideband drive.

\section{Conclusion and outlook}\label{sec:con_discussion}
We have proposed a protocol that significantly accelerates SNAP gates, surpassing the fundamental speed limit imposed by the bare dispersive shift by more than an order of magnitude. This acceleration is enabled by sideband engineering, which lifts the fundamental speed limit through enhancement of the effective dispersive shift. The resulting Floquet SNAP gate operates in the Floquet basis defined by the sideband drive, which requires modified frequencies and amplitudes.
Building on this framework, we have successfully extended optimal control techniques to the Floquet SNAP gate, achieving an additional speedup of more than an order of magnitude while maintaining high fidelity. Finally, through numerical simulations and analytical derivations, we have identified the dominant drive-induced decoherence mechanisms. Importantly, these effects do not compromise the fidelity gains enabled by Floquet acceleration.

The developed protocol is particularly useful in systems featuring ultra-high-coherence cavity modes~\cite{coh_1,coh_2,coh_3}, where dispersive shifts are typically kept small to minimize ancilla-induced cavity loss.  
During gate operation, the sideband drive is temporarily activated, enabling fast and high-fidelity SNAP operations.  
In the idling period, the sideband drive is turned off, restoring the cavity’s high coherence for quantum information storage.  
The accelerated operations could potentially be made fault-tolerant by employing two sideband drives simultaneously, enhancing both the dispersive shifts associated with the ancilla states $|e\rangle$ and $|f\rangle$ states, while making them equal~\cite{fault_tolerant_snap,fault_tolerant_2,Reinhold2020-ho}.
Beyond single-mode systems, our protocol also offers advantages in multimode architectures~\cite{Naik2017a,Chakram2020,Chakram2022,Alessandro, huang2025fastsidebandcontrolweakly}, where a single ancilla is coupled to multiple cavity modes.
In such setups, weak cavity\textendash ancilla coupling is helpful for reducing crosstalk, both directly through cross-Kerr interactions between cavity modes and indirectly via dispersive coupling between each cavity mode and the shared ancilla~\cite{you2024}.  
Our approach dynamically enhances the dispersive shift between the target mode and the ancilla to enable fast gate operation, while keeping both the cross-Kerr interaction and the dispersive shift of spectator modes minimal.  
Furthermore, our protocol can temporarily enhance the dispersive shift to accelerate parity mapping, a key step in quantum state tomography.

More broadly, enabling dynamic control over the dispersive shift enhances hardware adaptability across a variety of control protocols. For instance, the same hardware can support both the ECD gate, which is optimized for small dispersive shifts, and the SNAP gate, which operates more efficiently with larger dispersive shifts. 
Even more importantly, this tunability allows these gates to be applied sequentially within a single algorithm, unlocking hybrid control strategies that leverage the strengths of both protocols.
Since the efficiency of quantum algorithm decomposition can vary significantly with the control protocol employed, this tunability facilitates the development of flexible, protocol-agnostic hardware capable of supporting a wide range of quantum applications.

\begin{acknowledgments}
This material is based upon work supported by the U.S. Department of Energy, Office of Science, National Quantum Information Science Research Centers, Superconducting Quantum Materials and Systems Division (SQMS) under contract number DE-AC02-07CH11359.
\end{acknowledgments}

\appendix

\section{Detailed derivations of selected equations}\label{APP:DERIVATION}
In this section, we provide detailed derivations of key equations presented in the main text. We begin with a review of the dispersive transformation~\cite{blais_rmp}, which leads to the derivation of the dispersive Hamiltonian and the relevant sideband interactions. Next, we apply a Schrieffer–Wolff transformation (SWT)~\cite{cohen1998atom} to analyze the sideband interactions, resulting in a modified dispersive shift and  additional noise channels. 

\subsection{Effects of the dispersive interaction}\label{APP:DERIVATION_DISPERSIVE}
The system consists of a cavity coupled to an ancilla, which is driven by an external field. The static Hamiltonian is 
\begin{equation}
    \hat{H}_\text{JC} = \frac{\alpha}{2}\hat{q}^{\dag2}\hat{q}^2 
    + \omega_\text{q}\hat{q}^\dag \hat{q}
    + \omega_\text{c}\hat{a}^\dag \hat{a}
    + g(\hat{a} +\hat{a}^\dag)( \hat{q}^\dag +  \hat{q}),
\end{equation}
where \( \hat{a} \) and \( \hat{q} \) are the ladder operators for the cavity and ancilla, respectively, with corresponding frequencies \( \omega_\text{c} \) and \( \omega_\text{q} \). The interaction strength is denoted by \( g \).  
For a transmon ancilla~\cite{koch2007}, the anharmonicity is given by \( \alpha = -E_\text{J} \phi_\text{q}^4 / 2 \), where \( E_\text{J} \) is the Josephson energy and \( \phi_\text{q} \) represents the zero-point fluctuation.  
The total Hamiltonian includes an additional drive applied to the ancilla,
\begin{equation}
    \hat{H}_\text{tot} = \hat{H}_\text{JC} + \epsilon \cos(\omega_\text{d}t) (\hat{q}^\dag + \hat{q}), 
\end{equation}
where $\epsilon$ and $\omega_\text{d}$ characterize the drive amplitude and frequency. The Hamiltonian in the rotating frame, after applying the rotating-wave approximation (RWA) to both the coupling and drive terms, simplifies to  
\begin{equation*}
    \hat{H}_\text{r} = \frac{\alpha}{2}\hat{q}^{\dag2}\hat{q}^2 
    + g(\hat{a} \hat{q}^\dag e^{i\Delta t} + \hat{a}^\dag \hat{q} e^{-i\Delta t}) + \frac{\epsilon}{2}(\hat{q}^\dag e^{i\Delta_\text{q}t} + \hat{q} e^{-i\Delta_\text{q}t}),
\end{equation*}
with detunings  $\Delta=\omega_\text{q} - \omega_\text{c}$ and $\Delta_\text{q} = \omega_\text{q} - \omega_\text{d}$. 
For our purposes, the ancilla can be effectively described using its lowest four energy levels: \( |g\rangle \), \( |e\rangle \), \( |f\rangle \), and \( |h\rangle \). Accordingly, we truncate the Hilbert space of the ancilla and apply a unitary transformation to eliminate the Kerr nonlinearity. The resulting Hamiltonian takes the form  
\begin{equation*}
    \begin{aligned}
        \hat{H}_\text{r}' & = g\hat{a}^\dag (|g\rangle\langle e| + \sqrt{2} |e\rangle\langle f| e^{-i\alpha t} + \sqrt{3} |f\rangle \langle h| e^{-2i\alpha t}) e^{-i\Delta t}\\
    & + \frac{\epsilon}{2}(|g\rangle\langle e| + \sqrt{2} |e\rangle\langle f| e^{-i\alpha t} + \sqrt{3} |f\rangle \langle h| e^{-2i\alpha t}) e^{-i\Delta_\text{q} t} \\
    & + \text{h.c.}
    \end{aligned}
\end{equation*}

In the dispersive regime, where $g\ll\Delta$, we diagonalize the cavity\textendash ancilla interaction using a SWT, with generators obtained by solving 
\begin{equation}
    i\partial_t\hat{W}_j + g\hat{a}^\dag \sqrt{j+1} |j\rangle\langle j+1| e^{-i (j\alpha+\Delta) t} + \text{h.c.} = 0,
\end{equation}
where $j=g,e,f$ (or $j=0,1,2$ in numeric notation). 
The resulting generators are 
\begin{equation}
    \hat{W}_j  = \frac{\sqrt{j+1}g}{\Delta+j\alpha} e^{i[\Delta+j\alpha] t}\hat{a} |j+1\rangle \langle j| - \text{h.c.}
\end{equation}
To second order, applying the Baker–Campbell–Hausdorff formula yields the effective Hamiltonian 
\begin{equation*}
    \begin{aligned}
        \hat{H}_\text{eff} &= \hat{a}^\dag \hat{a} (\chi_g|g\rangle \langle g|  + \chi_e|e\rangle \langle e| + \chi_f|f\rangle \langle f|+ \chi_h|h\rangle \langle h|  ) \\
        &-\frac{\sqrt{2}}{2}\frac{\epsilon g\alpha}{\Delta(\Delta+\alpha)}
        \left[e^{-i(\Delta+\Delta_\text{q}+\alpha)t}\hat{a}^\dag |g\rangle \langle f| + \text{h.c.} \right]\\
        &-\frac{\sqrt{6}}{2}\frac{\epsilon g\alpha}{(\Delta+\alpha)(\Delta+2\alpha)}\left[e^{-i(\Delta+\Delta_\text{q}+3\alpha)t}\hat{a}^\dag |e\rangle \langle h|+ \text{h.c.} \right] \\ 
        & + \Big[\frac{\epsilon}{2}(|g\rangle\langle e| + \sqrt{2} |e\rangle\langle f| e^{-i\alpha t} + \sqrt{3} |f\rangle \langle h| e^{-2i\alpha t}) e^{-i\Delta_\text{q} t} \\
        & + \text{h.c.} \Big], 
    \end{aligned}
\end{equation*}
with dispersive shifts $\chi_g = -g^2/\Delta$, $\chi_e = 2g^2(\Delta+2\alpha)/\Delta/(\Delta+\alpha)$, and $\chi_f = -g^2 (\Delta-\alpha)/(\Delta+\alpha)/(\Delta+2\alpha)$. 
Within the $|g\rangle$\textendash$|e\rangle$ subspace, they lead to the well-known dispersive shift $\chi_0=\chi_e - \chi_g = 2g^2\alpha/\Delta/(\Delta+\alpha)$~\cite{blais_rmp}. Note that we have omitted the Lamb shifts, whose only effect is to renormalize the ancilla transition frequencies.
The second and third lines in $\hat{H}_\text{eff}$ describe the sideband interactions $|g,n+1\rangle \leftrightarrow |f,n\rangle $ and $|e,n+1\rangle \leftrightarrow |h,n\rangle $, respectively. 
In the limit $\Delta \gg \alpha$, the sideband rate reduces to the familiar expression from black-box quantization (BBQ)~\cite{fault_tolerant_snap},
\begin{equation*}
    \frac{\epsilon g\alpha}{\Delta(\Delta+\alpha)} \approx 
    \frac{\epsilon g\alpha}{(\Delta+\alpha)(\Delta+2\alpha)} \approx 
    \frac{\epsilon g\alpha}{\Delta^2} = -E_\text{J}\xi\phi_\text{c}\phi_\text{q}^3,
\end{equation*}
where we define the displacement as $\xi=\epsilon/(2\Delta)$, and the cavity participation factor as $\phi_\text{c}$. 

\begin{figure}[t]
    \centering
    \includegraphics[width=\columnwidth]{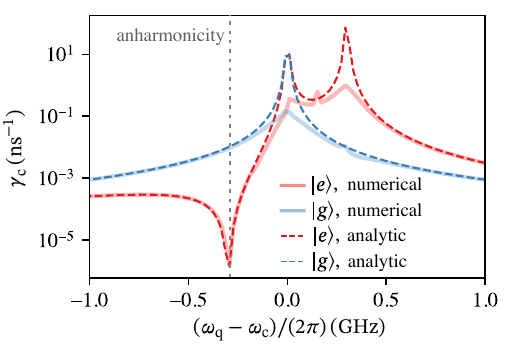}
    \caption{
    Inverse Purcell decay rate of a cavity coupled to a lossy ancilla.
    The blue and red curves represent the cases where the ancilla is in $|g\rangle$ and $|e\rangle$, respectively. The results from the Lindblad master equation (solid lines) show excellent agreement with the analytic expression (dashed lines). Notably, they both capture the vanishing decay rate when the detuning matches the anharmonicity of the ancilla (indicated by the gray dotted line). Parameters: $\omega_\text{q}/2\pi=6\,\text{GHz}$, $g/2\pi=30\,\text{MHz}$, $\alpha/2\pi=-300\,\text{MHz}$, $\gamma_\text{q}=(1\,\text{ns})^{-1}$.}
    \label{fig:FIG_APP_PURCELL} 
\end{figure}

The above derivation focuses on the effect of the dispersive transformation on the coherent dynamics. However, the transformation also modifies incoherent transitions and decoherence~\cite{dispersive_decoherence}. 
To leading order, an operator transforms as $\hat{\mathcal{O}} \to \hat{\mathcal{O}} + [\hat{W}, \hat{\mathcal{O}}]$. In the interaction picture, the ancilla decay and dephasing operators become
\begin{gather*}
    \begin{aligned}
        \hat{q}(t) & \to  \hat{q}(t)
    - \frac{g}{\Delta}\hat{a} |g\rangle \langle g| e^{-i\omega_\text{c}t}
    + \frac{g}{\Delta}\left( \frac{\alpha-\Delta}{\alpha+\Delta}\right)\hat{a} |e\rangle \langle e| e^{-i\omega_\text{c}t}\\
    &-\frac{\sqrt{2}g\alpha}{\Delta(\Delta+\alpha)} \hat{a}^\dag |g\rangle \langle f| e^{-i(2\omega_\text{q}+\alpha - \omega_\text{c})t}, 
    \end{aligned}\\
    \begin{aligned}
    \hat{q}^\dag\hat{q}(t) & \to  \hat{q}^\dag\hat{q}(t) 
    - \frac{g}{\Delta}\hat{a}^\dag |g\rangle \langle e| e^{-i\Delta t}
    - \frac{g}{\Delta}\hat{a} |e\rangle \langle g| e^{i\Delta t}\\
    &- \frac{\sqrt{2}g}{\Delta+\alpha}\hat{a}^\dag |e\rangle \langle f| e^{-i(\Delta +\alpha)t}
    - \frac{\sqrt{2}g}{\Delta+\alpha}\hat{a} |f\rangle \langle e| e^{i(\Delta +\alpha)t}.
    \end{aligned}
\end{gather*}
The exponents in these expressions indicate the frequencies at which the noise spectral density is evaluated—that is, the frequency components where noise has a significant impact.
For example, the frequency $\omega_\text{c}$ associated with the term $\hat{a}|g\rangle\langle g|$ corresponds to energy decay from the cavity to the bath, while the frequency $\Delta + \alpha$ associated with $\hat{a}^\dag |e\rangle \langle f|$ describes a cavity excitation accompanied by ancilla decay from $|f\rangle$ to $|e\rangle$, with the frequency mismatch compensated by the bath.
Notably, the rate of ancilla-induced cavity decay (inverse Purcell decay) depends on the state of the ancilla. 
When the ancilla is in the ground state, the decay rate is $(g^2/\Delta^2 )\gamma_\text{q}$, with $\gamma_\text{q}$ the intrinsic ancilla decay rate;
When the ancilla is in the first excited state, the induced decay rate is modulated by an additional factor $(\alpha-\Delta)^2/(\alpha+\Delta)^2$. A particularly interesting case arises when $\Delta = \alpha$, where this modulation factor vanishes, leading to a suppression of cavity decay.
We emphasize that this condition is distinct from the straddling regime, which corresponds to $\Delta \approx -\alpha$.
We confirm this behavior through numerical simulations of the master equation, as evidenced by the red dip in Fig.~\ref{fig:FIG_APP_PURCELL}. Despite the overall agreement, noticeable deviations arise between the numerical diagonalization and the analytical expression near $\Delta \approx 0$ and $\Delta \approx -\alpha$, where the dispersive approximation breaks down.

\subsection{Effects of the sideband interaction}\label{APP:DERIVATION_SIDEBAND}
The sideband interactions derived above take the general form of 
\begin{equation}
   \hat{H}_\text{sb} = \frac{\Omega_j}{2}(\hat{a}^\dag |j\rangle \langle j+2|e^{-i\Delta_j t} + \text{h.c.}),
\end{equation}
where the indices $j=g,e$ correspond to the sideband transitions $|g,n+1\rangle \leftrightarrow |f,n\rangle $ and $|e,n+1\rangle \leftrightarrow |h,n\rangle $, respectively. 
The sideband rates are given by $\Omega_g = -\sqrt{2}\epsilon g\alpha/\Delta/(\Delta+\alpha)$ and $\Omega_e = -\sqrt{6}\epsilon g\alpha/(\Delta+\alpha)/(\Delta+2\alpha)$, with detunings $\Delta_g = \Delta+\Delta_\text{q}+\alpha$ and $\Delta_e = \Delta+\Delta_\text{q}+3\alpha$. 
In the regime where $\Delta_j \gg \Omega_j$, we perform a SWT to diagonalize the sideband Hamiltonian. The generators of the transformations are 
\begin{equation}
    \hat{G}_j  =  \frac{\Omega_i}{2\Delta_i}e^{i\Delta_j t}\hat{a}|j+2\rangle \langle j| - \text{h.c.},
\end{equation}
which leads to a driven analog to the the dispersive Hamiltonian~\cite{fault_tolerant_snap}
\begin{equation*}
    \hat{H}_\text{sb}' = \sum_{j=g,e} \left(-\frac{\Omega_j^2}{4\Delta_j} \hat{a}^\dag \hat{a} |j\rangle \langle j| + \frac{\Omega_j^2}{4\Delta_j} \hat{a}^\dag \hat{a} |j+2\rangle \langle j+2|\right).
\end{equation*}
Depending on the drive frequency, one of the sideband interaction dominates, leading to a modified dispersive shift
\begin{align}
    \chi_\text{d}^{g,n+1\leftrightarrow f,n} &= \chi_0 +  \frac{\Omega_g^2}{4\Delta_g}, \\
    \chi_\text{d}^{e,n+1\leftrightarrow h,n} &= \chi_0 -
    \frac{\Omega_e^2}{4\Delta_e}.
\end{align}

Next, we examine how the ancilla operators transform under the sideband drive, which affects both the response to external drives during gate operations and the coupling to the bath, leading to decoherence. 
To leading order, the ancilla decay and dephasing operators transform as  
\begin{equation*}
    \begin{aligned}
        \hat{q}(t) \to & \hat{q}(t)
    +  \frac{\sqrt{3}\Omega_g}{2\Delta_g}\hat{a}^\dag|g\rangle\langle h|e^{-i(\Delta_g+\omega_\text{q}+2\alpha)t}\\
    & - \frac{\Omega_g}{2\Delta_g}\hat{a}|f\rangle\langle e|e^{i(\Delta_g-\omega_\text{q})t}  - \frac{\Omega_e}{2\Delta_e}\hat{a}^\dag|g\rangle\langle h|e^{-i(\Delta_e+\omega_\text{q})t} \\
    & + \frac{\sqrt{2}\Omega_g}{2\Delta_g}\hat{a}|e\rangle\langle g|e^{i(\Delta_g-\omega_\text{q}-\alpha)t}\\
    &    
   + \frac{\sqrt{3}\Omega_e}{2\Delta_e}\hat{a}|f\rangle\langle e|e^{i(\Delta_e-\omega_\text{q}-2\alpha)t}\\
    & - \frac{\sqrt{2}\Omega_e}{2\Delta_e}\hat{a}|h\rangle\langle f|e^{-i(\Delta_e-\omega_\text{q}-\alpha)t}, 
    \end{aligned}
\end{equation*}
\begin{equation*}
    \begin{aligned}
       \hat{q}^\dag\hat{q}(t) \to & \hat{q}^\dag\hat{q}(t)
    +  \frac{\Omega_g}{\Delta_g}\hat{a}^\dag|g\rangle\langle f|e^{-i\Delta_g t}
    +  \frac{\Omega_g}{\Delta_g}\hat{a}|f\rangle\langle g|e^{i\Delta_g t}\\
    & +  \frac{\Omega_e}{\Delta_e}\hat{a}^\dag|e\rangle\langle h|e^{-i\Delta_e t}
     + \frac{\Omega_e}{\Delta_e}\hat{a}|h\rangle\langle e|e^{i\Delta_e t}.
    \end{aligned}
\end{equation*}
We have neglected cross terms arising from both the dispersive and sideband transformations, as these are higher-order corrections.
Since the computation states of the ancilla consists of $|g,e\rangle$, only terms involving $\langle g,e|$ are relevant for incoherent transitions.

\section{Floquet\textendash Markov equation with pure dephasing}\label{app:derivation_FM}
In this appendix, we present a detailed derivation of the Floquet–Markov equation governing the decoherence of a driven system coupled to a thermal bath, as employed in Sec.~\ref{sec:noise}. While incoherent transitions have been extensively studied in the context of driven open quantum systems~\cite{hanggi1991,PhysRevApplied.11.024003}, the role of pure dephasing has been largely neglected, despite its critical impact on decoherence in many practical settings. Only recently has this issue begun to receive focused attention, as in the work of Zhang et al.~\cite{yaxing}. Here, we go beyond previous treatments by rigorously deriving the Floquet–Markov equation and demonstrating that pure dephasing terms emerge systematically and unambiguously when the rotating-wave approximation is applied with proper care. Our approach mainly follows the formalism outlined in the review by Grifoni and Hänggi~\cite{hanggi1991}. Finally, we validate our results by comparing numerical simulations with the derived analytical expressions. 

\vspace{1em}

\subsection{Master equation from second-order perturbation theory}
Consider a quantum system S with a time-periodic Hamiltonian, \( \hat{H}_\text{S}(t + T) = \hat{H}_\text{S}(t) \), interacting with a bath B. The total Hamiltonian is given by:
$\hat{H}_\text{S}(t) + \hat{H}_\text{B} + \hat{H}_\text{SB}$, where $\hat{H}_\text{B}$ is the bath Hamiltonian, and $\hat{H}_\text{SB}$ describes the system\textendash bath interaction. 
Following the common Caldeira\textendash Leggett approach~\cite{CALDEIRA1983374}, we model the bath as an ensemble of harmonic oscillators with the coupling Hamiltonian  
$\hat{H}_\text{SB} = \hat{A} \sum_{\nu} g_\nu \hat{x}_\nu$,
where $\hat{A}$ is the system operator, $\hat{x}_\nu$ is the position operator of mode $\nu$, and $g_\nu$ represents the coupling strength between them. 
Applying second-order perturbation theory to the coupling term, the density matrix in the lab frame obeys the master equation 
\begin{equation*}
\begin{aligned}
    \dot{\rho}_\text{S}(t) =& -i \left[ \hat{H}_\text{S}(t), \rho_\text{S}(t) \right] 
     -i \, \text{Tr}_\text{B} \left[ \hat{H}_\text{SB}, \rho_\text{S}(t) \right] \\
    &- \int_{0}^{\infty} d\tau \, \text{Tr}_\text{B} \left[ \hat{H}_\text{SB}, \left[\tilde{H}_\text{SB}(t - \tau, t), \rho_\text{B} \otimes \rho_\text{S}(t) \right]\right].
\end{aligned}
\end{equation*}
The final integral captures the accumulated influence of past system\textendash bath interactions at time $t - \tau$, which irreversibly affect the system's dynamics at the present time $t$.
The tilde denotes an operator in the interaction picture, defined as 
$\tilde{O}(t, t') = \hat{U}_0^{\dagger}(t, t') \hat{O} \hat{U}_0(t, t')$,
where the unitary operator is give by
\begin{equation}
    \hat{U}_0(t, t') = \mathcal{T} \exp \left( -\frac{i}{\hbar} \int_{t'}^{t} dt'' \left[ \hat{H}_\text{S}(t'') + \hat{H}_\text{B} \right] \right),
\end{equation}
with $\mathcal{T}$ denoting the time-ordering operator.
Substituting the explicit form of the coupling Hamiltonian $\hat{H}_\text{SB}$ into the last term of the master equation, we obtain the expression 
\begin{widetext}
\begin{equation}\label{eq:master_eqn}
    \begin{aligned}
        I & = \int_{0}^{\infty} d\tau \, \text{Tr}_\text{B} \left[ \hat{H}_\text{SB}, \tilde{H}_\text{SB}(t - \tau, t), \rho_\text{B} \otimes \rho_\text{S}(t) \right] \\
        & = \sum_{\nu,\mu}g_\nu g_\mu \int_0^\infty \,\mathrm{d}\tau\left\{
        [\hat{A},\tilde{A}(t-\tau, t)\rho_\text{S}(t)] \,\text{Tr}_\text{B}\left[\hat{x}_\nu\tilde{x}_\mu(t-\tau, t) \rho_\text{B}\right]
        - [\hat{A},\rho_\text{S}(t)\tilde{A}(t-\tau, t)] \,\text{Tr}_\text{B}\left[\tilde{x}_\mu(t-\tau, t) \hat{x}_\nu\rho_\text{B}\right]
        \right\} \\
        & = \sum_{\nu} g_\nu^2 \int_0^\infty\,\mathrm{d}\tau \left\{
        C_\nu^\text{sym}(\tau)[\hat{A},[\tilde{A}(t-\tau, t), \rho_\text{S}(t)]] 
        +   iC_\nu^\text{anti}(\tau)[\hat{A},\{\tilde{A}(t-\tau, t), \rho_\text{S}(t)\}]\right\}  \\ 
        & = \sum_{\nu} \frac{g_\nu^2}{2m_\nu\omega_\nu} \int_0^\infty\,\mathrm{d}\tau \left\{
         \coth{(\beta\omega_\nu/2)} \cos(\omega_\nu \tau)[\hat{A},[\tilde{A}(t-\tau, t), \rho_\text{S}(t)]] 
        -   i \sin(\omega_\nu \tau)[\hat{A},\{\tilde{A}(t-\tau, t), \rho_\text{S}(t)\}]\right\}\\
         &=\int_0^\infty \mathrm{d}\omega\, J(\omega)  \int_0^\infty\,\mathrm{d}\tau \left\{[e^{i\omega\tau} n_\omega+ e^{-i\omega\tau} (n_\omega+1)] [\hat{A}\tilde{A}(t-\tau, t)\rho_\text{S}(t) - \tilde{A}(t-\tau, t)\rho_\text{S}(t)\hat{A}(t)] + \text{h.c.}
     \right\}.
    \end{aligned}
\end{equation}
\end{widetext}
In deriving the second and third equalities of the above equations, we use the common definitions of the symmetric and anti-symmetric correlation functions~\cite{correlation} for mode \( \nu \), assuming no correlations between different modes,
\begin{align}
    C_\nu^{\text{sym}}(t) &= \frac{1}{2} \langle \{ \hat{x}_\nu(t), \hat{x}_\nu(0) \} \rangle,  \\
    C_\nu^{\text{anti}}(t) &= \frac{1}{2i} \langle [ \hat{x}_\nu(t), \hat{x}_\nu(0) ] \rangle.
\end{align}
Taking the exact form of the position operator of a harmonic oscillator
$\hat{x}_\nu(t) = \sqrt{\hbar/2 m_\nu \omega_\nu} \left( \hat{a}_\nu e^{-i \omega_\nu t} + \hat{a}_\nu^\dagger e^{i \omega_\nu t} \right)$, and assuming the bath is in thermal equilibrium, the correlation functions take the form
\begin{align}
    C_\nu^{\text{sym}}(t) &= \left( \frac{\hbar}{2 m_\nu \omega_\nu} \right) \left[ \coth{(\beta\omega_\nu/2)} \cos(\omega_\nu t) \right],  \\
    C_\nu^{\text{anti}}(t) &= - \left( \frac{\hbar}{2 m_\nu \omega_\nu} \right)  \sin(\omega_\nu t).
\end{align}
In cQED, the bath is often modeled as a transmission line~\cite{As2003}. In this setting, the prefactors in the above expressions are replaced by $\hbar Z_m / 2$, where $Z_m$ denotes the impedance of mode $m$.
Finally, in the last equality of Eq.~\eqref{eq:master_eqn}, we convert the discrete sum over modes into an integral, and introduce the spectral density of the bath oscillators~\cite{RevModPhys.59.1}
\begin{equation}
    J(\omega) = \sum_\nu \frac{g_\nu^2}{2m_\nu \omega_\nu} \delta(\omega - \omega_\nu).
\end{equation}
Note that different conventions exist for defining the spectral density. Here, we adopt the definition without an explicit factor of $\pi$ to ensure consistency with the results in Ref.~\onlinecite{hanggi1991}.

\subsection{Master equation in the Floquet basis}
To facilitate numerical simulations, we express the master equation in an algebraic form within a specific basis. 
Given the presence of a periodic drive, it is natural to employ a basis such as Floquet modes or Floquet states. Floquet modes can be defined within any Brillouin zone, leading to different wavefunctions and quasienergies. However, Floquet modes corresponding to various Brillouin zones represent the same Floquet state. Therefore, we adopt Floquet states as the basis for our formulation. 
Specifically, the matrix elements of an operator $\hat{A}$ in the basis of Floquet states, $|\psi_{a,b}(t)\rangle = |\phi_{a,b}(t)\rangle e^{-i\epsilon_{a,b} t}$, take the form  
\begin{equation}
    \langle \psi_a(t) | \hat{A} | \psi_b(t) \rangle = \langle \phi_a(t) | \hat{A} | \phi_b(t) \rangle e^{i(\epsilon_a - \epsilon_b)t}.
\end{equation}
Since the Floquet modes $|\phi_{a,b}(t)\rangle$ are periodic in time with frequency $\Omega=2\pi/T$, the matrix elements can be expanded in a Fourier series as
\begin{equation}
    \langle \phi_a(t) | \hat{A} | \phi_b(t) \rangle \equiv A_{a,b}(t) =  \sum_k e^{ik\Omega t} A_{ab,k},
\end{equation}
where the Fourier coefficients are given by
\begin{equation}
    A_{ab,k}=  \frac{1}{T} \int_0^T \mathrm{d}\tau \,e^{-i k \Omega \tau} \langle \phi_a(\tau) | \hat{A} | \phi_b(\tau) \rangle. 
\end{equation}
The matrix elements in the Floquet states basis simplify to
\begin{equation}
    \langle \psi_a(t) | \hat{A} | \psi_b(t) \rangle  = \sum_k e^{i\Delta_{ab,k}t}A_{ab,k},
\end{equation}
with the energy difference $\Delta_{ab,k} = \epsilon_a - \epsilon_b + k\Omega$. 
Physically, $A_{ab,k}$ represents the transition between states in different Brillouin zones, assisted by the absorption or emission of $k$ photons. 

To simplify the master equation in the Floquet basis, we introduce the following quantities, 
\begin{align}
    R_{ab}(t,\tau) =& \langle \psi_a(t)| \hat{A}\tilde{A}(t-\tau, t)\rho_\text{S}(t)| \psi_b(t)\rangle \notag \\ 
    &- \langle \psi_a(t)| \tilde{A}(t-\tau, t)\rho_\text{S}(t)\hat{A}(t) | \psi_b(t)\rangle \notag \\
    =& \sum_{a',b'} A^*_{a'a}(t) A_{a'b'}(t-\tau)\rho_{b'b}(t) \\
    &- A_{aa'}(t-\tau)\rho_{a'b'}(t)A^*_{bb'}(t), \notag
\end{align}
\begin{align}
    S_{ab}(t,\tau) =& \langle \psi_a(t)| \rho_\text{S}(t)\tilde{A}(t-\tau, t)\hat{A}| \psi_b(t)\rangle \notag \\
    &-  \langle \psi_a(t)|\rho_\text{S}(t)\hat{A}(t)\tilde{A}(t-\tau, t) | \psi_b(t)\rangle \notag \notag\\
    =& \sum_{a',b'} \rho_{aa'}(t) A_{a'b'}(t-\tau) A^*_{bb'}(t) \\
    &- A^*_{a'a}(t) \rho_{a'b'}(t) A_{b'b}(t-\tau). \notag
\end{align}
Here, the operators in the interacting picture are obtained using the relation
\begin{equation}
\begin{aligned}
    &\langle \psi_a(t)| \tilde{A}(t-\tau, t) | \psi_b(t)\rangle \\
    &= \langle \psi_a(t)|\hat{U}_0^\dag(t-\tau, t) \hat{A} \hat{U}_0(t-\tau, t)| \psi_b(t)\rangle \\
    &= \langle \psi_a(t-\tau)| \hat{A} | \psi_b(t-\tau)\rangle.
\end{aligned}
\end{equation}
With the definition of $R_{ab}$ and $S_{ab}$, the interaction term in the master equation simplifies to the following form
\begin{equation*}
    \begin{aligned}
     I_{ab} = \int_0^\infty \mathrm{d}\omega\, J(\omega)  \int_0^\infty\,\mathrm{d}\tau 
     \bigg\{
        &e^{i\omega\tau} n(\omega) R_{ab}(t-\tau) \\
        + &e^{-i\omega\tau} [n(\omega)+1] R_{ab}(t-\tau)\\
        + &e^{-i\omega\tau} n(\omega) S_{ab}(t-\tau) \\
        + &e^{i\omega\tau} [n(\omega)+1] S_{ab}(t-\tau)
     \bigg\}.
    \end{aligned}
\end{equation*}
In the following, we provide the derivation of the first term as an example and omit the remaining three, as they can be derived analogously. 
Specifically, we obtain
\begin{widetext}
\begin{equation}\label{eq:r_ab}
    \begin{aligned}
        I_{ab}^{(1)}& = \int_0^\infty \mathrm{d}\omega\, J(\omega)  \int_0^\infty\,\mathrm{d}\tau 
        e^{i\omega\tau} n(\omega) R_{ab}(t-\tau) \\
        &= \pi\sum_{a',b'} \sum_{m,n} e^{-i(\Delta_{a'a,m} - \Delta_{a'b',n})t}A^*_{a'a,m} A_{a'b',n}\rho_{b'b}(t) J(\Delta_{a'b',n}) n(\Delta_{a'b',n}) \\
        & \qquad\qquad - e^{i(\Delta_{aa',m} - \Delta_{bb',n})t}A_{aa',m} \rho_{a'b'}(t) A^*_{bb',n} J(\Delta_{aa',m})n(\Delta_{aa',m}),
    \end{aligned}
\end{equation}
\end{widetext}
where we have used the Fourier series expansion of the matrix elements and the approximation $\int_0^\infty \mathrm{d}\tau e^{i\omega\tau} \approx \pi \delta(\omega) $. 

\vspace{-1ex}

\subsection{Rotating-wave approximation}
To further simplify the results, we apply the rotating-wave approximation (RWA), neglecting the terms that oscillate at frequencies much higher than the relevant timescale of the system dynamics. 
This assumption can break down if the system's transition energies exhibit degeneracies either within a given Brillouin zone or across different zones. 

\subsubsection{Case I: Off-diagonal contribution, $a\neq b$}
For $a\neq b$, corresponding to dephasing dynamics, the exponent in the first term of $I_{ab}^{(1)}$  vanishes when $b'=a$ and $m=n$, reducing the expression to
\begin{equation}
    \pi\sum_{a'} \sum_{m} |A_{a'a,m}|^2 \rho_{ab}(t) J(\Delta_{a'a,m}) n(\Delta_{a'a,m}).
\end{equation}
Similarly, in the second term, only the contributions for $a=a'$, $b=b'$, and $m=n$ survive under the RWA, leading to
\begin{equation}\label{eq:FM_dephasing}
    -\pi \sum_{m} A_{aa,m}A_{bb,m}^*\rho_{ab}(t) J(\Delta_{aa,m}) n(\Delta_{aa,m}).
\end{equation}
Notably, this term is often neglected in the literature, resulting in an incomplete characterization of the pure dephasing process. A more detailed discussion of its implications will be provided later in this section.

Applying similar derivations to the remaining three terms in $I_{ab}$, we obtain
\begin{widetext}
\begin{equation}
    \begin{aligned}
        I_{ab} & =  \pi\sum _{\nu\neq a} \sum _{k} |A_{\nu a,k}|^2 \rho_{ab}(t) \{J(\Delta_{\nu a,k}) n(\Delta_{\nu a,k}) + J(-\Delta_{\nu a,k}) [n(-\Delta_{\nu a,k})+1]\} \\
       & + \pi\sum _{\nu\neq b} \sum _{k} |A_{b\nu,k}|^2 \rho_{ab}(t) \{J(\Delta_{b\nu,k}) [n(\Delta_{b\nu,k}) +1] + J(-\Delta_{b\nu,k}) n(-\Delta_{b\nu,k}) \}\\
       & + \pi\sum_k |A_{aa,k}-A_{bb,k}|^2 \left\{J(\Delta_k)n(\Delta_k) +J(-\Delta_k)[n(-\Delta_k)+1]\right\}.
    \end{aligned}
\end{equation}
\end{widetext}
To align with the notation in Ref.~\onlinecite{hanggi1991}, we redefine the indices as $a'\to\nu, m\to k$, and introduce the notation $\Delta_k = k\Omega$. 
Since the spectral density $J(\omega)$ is defined only for positive arguments, we introduce a Heaviside function to enforce this condition. 
Additionally, we use the relation $-\Delta_{\nu a,k} = \Delta_{a\nu,-k}$ and $|A_{\nu a,k}| = |A_{a\nu,-k}|$. 
The master equation in the interacting picture then takes the form
\begin{widetext}
\begin{equation}
    \begin{aligned}
        \dot{\rho}_{ab}(t) = -\rho_{ab}(t) \pi \sum _{k} &\sum _{\nu\neq a} |A_{\nu a,k}|^2  \Theta(\Delta_{\nu a,k})J(\Delta_{\nu a,k}) n(\Delta_{\nu a,k})  
        +  |A_{a\nu,-k}|^2  \Theta(\Delta_{a\nu,-k})J(\Delta_{a\nu,-k}) [n(\Delta_{a\nu,-k})+1] \\
        + &\sum_{\nu \neq b}|A_{\nu b,k}|^2  \Theta(\Delta_{\nu b,k})J(\Delta_{\nu b,k}) n(\Delta_{\nu b,k})
      +  |A_{b\nu,-k}|^2  \Theta(\Delta_{b\nu,-k})J(\Delta_{b\nu,-k}) [n(\Delta_{b\nu,-k})+1] \\
       + & |A_{aa,k}-A_{bb,k}|^2 J(\Delta_k)[2n(\Delta_k)+1]. 
    \end{aligned}
\end{equation}
\end{widetext}

\noindent We further simplify the expression by introducing the following definitions
\begin{gather*}
        \gamma_{\alpha\beta, k} = 2\pi \Theta(\Delta_{\alpha\beta,k})J(\Delta_{\alpha\beta,k}) |A_{\alpha\beta,k}|^2, \\
        V_{\beta\alpha} = \sum_k \gamma_{\alpha\beta,k} n (\Delta_{\alpha\beta,k}) 
        + \gamma_{\beta\alpha,-k} [n (\Delta_{\beta\alpha,-k})+1],  \\
        W_{\alpha\beta} = 2\pi \sum_{k} \Theta(\Delta_k) |A_{\alpha\alpha,k}-A_{\beta\beta,k}|^2 J(\Delta_k)[2n(\Delta_k)+1].
\end{gather*}
With these definitions, the master equation simplifies to
\begin{equation}
    \dot{\rho}_{ab}(t) = - \dfrac{1}{2}\rho_{ab}(t) \Big(\sum _{\nu\neq a} V_{a\nu} + \sum_{\nu\neq b} V_{b\nu} + W_{ab}\Big).
\end{equation}
Physically, the first two terms describe the dephasing associated with transitions from states $a,b$ to $\nu$, while the last term characterizes pure dephasing. 

\subsubsection{Case II: Diagonal contribution, $a = b$}
For $a=b$, corresponding to dissipation dynamics, applying the RWA to Eq.~\eqref{eq:r_ab} results in
\begin{equation*}
\begin{aligned}
    I_{aa}^{(1)} = \pi\sum_{a'} \sum_{m} |A_{a'a,m}|^2 \rho_{aa}(t) J(\Delta_{a'a,m}) n(\Delta_{a'a,m}) \\
    + |A_{aa',m}|^2 \rho_{aa}(t) J(\Delta_{aa',m}) n(\Delta_{aa',m}).
\end{aligned}
\end{equation*}
Applying similar derivations to the remaining three terms in $I_{aa}$, the diagonal contribution to the master equation becomes
\begin{widetext}
    \begin{equation}
    \begin{aligned}
        \dot{\rho}_{aa}(t) = - 2\pi\sum _{\nu,k} 
        & \rho_{aa}(t)\{|A_{\nu a,k}|^2  \Theta(\Delta_{\nu a,k})J(\Delta_{\nu a,k}) n(\Delta_{\nu a,k}) + |A_{a\nu,-k}|^2  \Theta(\Delta_{a\nu,-k})J(\Delta_{a\nu,-k}) [n(\Delta_{a\nu,-k})+1]\}\\
        -& \rho_{\nu\nu}(t)\{|A_{a\nu,k}|^2  \Theta(\Delta_{a\nu,k})J(\Delta_{a\nu,k}) n(\Delta_{a\nu,k}) + |A_{\nu a,-k}|^2  \Theta(\Delta_{\nu a,-k})J(\Delta_{\nu a,-k}) [n(\Delta_{\nu a,-k})+1] \}.
    \end{aligned}
\end{equation}
\end{widetext}
Here, the terms associated with $n$ and $n+1$ correspond to absorption and emission processes, respectively.
Using the notations introduced earlier, the master equation simplifies to
\begin{equation}
    \dot{\rho}_{aa}(t) =  \sum_\nu \rho_{\nu\nu}(t) V_{\nu a} - \rho_{aa}(t)  V_{a\nu}. 
\end{equation}
Intuitively, the terms on the right-hand side describe all transitions from states $|\nu\rangle$ to the state $|a\rangle$, with associated rates $V_{\nu a}$, and vice versa.

\subsection{Numerical simulation of driven decoherence in a qubit}
\begin{figure}[t]
    \centering
    \includegraphics[width=\columnwidth]{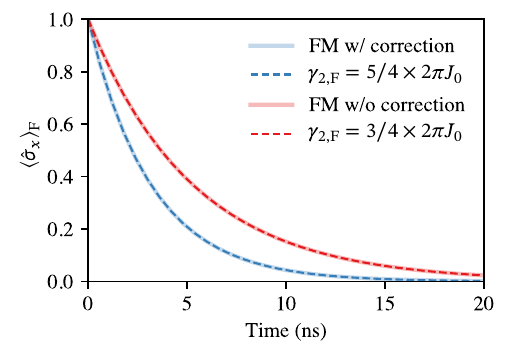}
    \caption{
    Numerical simulation of the driven qubit decoherence.
    The blue and red curves depict the Ramsey experiment in the Floquet basis, simulated using the Floquet\textendash Markov (FM) equation with and without the correction in Eq.~\eqref{eq:FM_dephasing}, respectively. 
    Analytical results with different dephasing rates are represented by dashed lines,  closely matching the corresponding simulations. Parameters: $\omega_\text{q}/2\pi = \omega_\text{d}/2\pi = 1 \,\text{GHz}$, $A/2\pi=0.1$ GHz, $J_0 = 0.04/2\pi$.}
    \label{fig:FIG_APP_FM} 
\end{figure}
To demonstrate the significance of Eq.~\eqref{eq:FM_dephasing}, we perform a numerical simulation of qubit decoherence under periodic drive. The system Hamiltonian is given by $\hat{H}_\text{q} = \frac{1}{2}\omega_\text{q}\hat{\sigma}_z + \frac{1}{2}A\cos(\omega_\text{d}t)\hat{\sigma}_x$, where the drive frequency $\omega_\text{d}$ is resonant with the qubit transition frequency $\omega_\text{q}$. 
For simplicity, we consider identical constant spectral densities $J(\omega \geq 0)= J_0$ for both the $\hat{\sigma}_x$ and $\hat{\sigma}_z$ couplings to the bath. 
The qubit is initialized in an equal superposition of the Floquet eigenstates, $|\psi(0)\rangle = (|1\rangle_\text{F} + |0\rangle_\text{F} )/\sqrt{2}$, and we calculate the expectation value of the Pauli operator $\hat{\sigma}_x$ in the Floquet basis. 
The results with and without the correction in Eq.~\eqref{eq:FM_dephasing} are shown in Fig.~\ref{fig:FIG_APP_FM} as blue and red curves, respectively. 
For a resonant drive, the analytical expression for the dephasing rate is $\gamma_{2,\text{F}} = 3\gamma_1/4 + \gamma_\nu/2$~\cite{Ithier2005,Yan2013}. In the case of constant spectral density, we have $\gamma_1 = \gamma_\nu = 2\pi J_0$. 
As expected, the analytical result (blue dashed lines) exactly matches the Floquet–Markov simulation when the correction term is included, whereas significant deviations appear when the correction is omitted. Specifically, the latter simulation matches the dephasing rate $3\gamma_1/4$ (red dashed lines), confirming that pure dephasing is not properly accounted for when the correction term is neglected.

\section{Effects of ramp time on adiabatic mapping}\label{APP:RAMP}

\begin{figure}[t]
\centering
\includegraphics[width=\columnwidth]{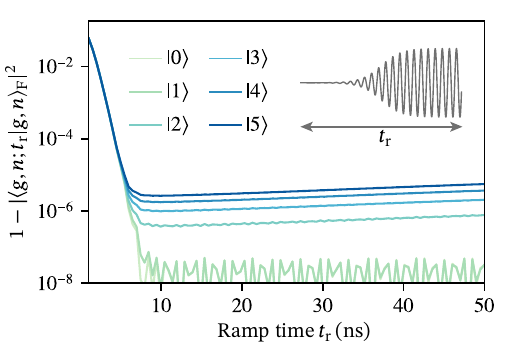}    
\caption{
Adiabatic mapping between the lab basis and the Floquet basis.
The overlap between the state $|g,n; t_\text{r}\rangle$ evolved under the ramp and the target Floquet state is shown as a function of the ramp time. 
Parameters: $\omega_\text{q}/2\pi=6.4$ GHz, $\omega_\text{c}/2\pi=4.5$ GHz, $\alpha/2\pi=-230$ MHz, $\chi_0/2\pi=0.14$ MHz, $\epsilon/2\pi = 0.8$ GHz, $\omega_\text{d}/2\pi = 7.56$ GHz.  
}
\label{fig:FIG_APP_RAMP}
\end{figure}

In Sec.~\ref{sec:floquet_snap}, we employ both ramp-up and ramp-down pulses to adiabatically map between states in the lab basis and the Floquet basis.  
To achieve high-fidelity mapping, the ramp must be sufficiently smooth such that its duration is significantly longer than the timescale set by the energy gap between the relevant Floquet states (i.e., \( |g,n\rangle_\text{F} \) and \( |e,n\rangle_\text{F} \))~\cite{adiabatic}.  
In Fig.~\ref{fig:FIG_APP_RAMP}, we characterize the fidelity of this mapping by computing the overlap between the state \( |g,n; t_\text{r}\rangle \) evolved under the ramp and the target Floquet state at the end of the ramp.  
The results are plotted as a function of the ramp duration.  
As expected, for short ramp times ($<10$ ns), the infidelity decreases sharply.
However, for longer ramp times, the infidelity saturates and exhibits a slight increase.
This behavior can be attributed to leakage into weakly coupled, highly excited folded states enabled by absorption of drive photons~\cite{adiabatic2}. Unlike static avoided crossings, those induced by photon absorption must be traversed diabatically to ensure that the system remains within the same Brillouin zone.
Therefore, there exists an optimal regime, where the ramp is slow enough to follow the standard crossings, yet fast enough to pass the photon-absorption crossings. 
In our implementation, we choose a ramp time of \( t_\text{r} = 10 \) ns, which lies within this optimal regime and ensures high-fidelity basis mapping.

\section{Hamiltonian implemented in numerical simulation}\label{APP:HAM}
This section details the system Hamiltonian used in full numerical simulations.
The full Hamiltonian, expressed in the black-box quantization (BBQ) framework~\cite{bbq}, is given by:
\begin{equation*}
    \hat{H}_\text{BBQ} = \tilde{\omega}_\text{c}\hat{c}^\dag\hat{c}  + \tilde{\omega}_\text{q}\hat{q}^\dag\hat{q}
    - E_\text{J}\cos_\text{NL}[\phi_\text{q}(\hat{q}^\dag + \hat{q}) + \phi_\text{c}(\hat{c}^\dag + \hat{c})],
\end{equation*}
where $\phi_\text{q}$ and $\phi_\text{c}$ denote the participation factors of the ancilla and cavity modes, respectively. The nonlinear cosine potential is expanded as  $\cos_\text{NL}(x) = \cos(x) - 1 + x^2/2$. 
Notably, we retain all fourth-order nonlinear terms, including fast-rotating ones, along with higher-order nonlinear terms beyond fourth order to ensure an accurate description of the system dynamics and capture all relevant interactions.
For driven system simulations, the truncation levels for the cavity and ancilla in the ladder basis are 12 and 20, respectively, to ensure numerical convergence.
The parameters used in the Hamiltonian are as follows: $\tilde{\omega}_\text{c}/2\pi=4.5$ GHz,  $\tilde{\omega}_\text{q}/2\pi=6.6$ GHz, $E_\text{J}/2\pi=26$ GHz, $\phi_\text{c}=0.0053$, $\phi_\text{q}=0.357$. 
Solving for the eigenvalues of the Hamiltonian, we obtain the following system parameters: dressed ancilla frequency $\omega_\text{q}/2\pi=6.4$ GHz, dressed cavity frequency $\omega_\text{c}/2\pi=4.5$ GHz, ancilla anharmonicity $\alpha/2\pi=-230$ MHz, dispersive shift $\chi_0/2\pi=0.14$ MHz. 

\section{Open-system gate fidelity}\label{APP:GATE_FIDELITY}

\begin{figure}[t]
    \centering
    \includegraphics[width=\columnwidth]{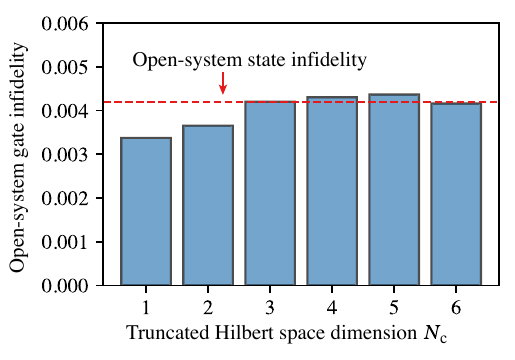}
    \caption{
    Open-system gate infidelity of the 1500-ns QOC Floquet SNAP gate from Fig.~\ref{fig:FIG_QOC_FLOQUET_SNAP}.
The horizontal axis represents the truncated cavity Hilbert space dimension $N_\text{c}$.
For reference, the open-system state infidelity from Fig.~\ref{fig:FIG_QOC_FLOQUET_SNAP} is shown as a dashed line.}
    \label{fig:FIG_APP_OPEN_FIDELITY} 
\end{figure}

In Sec.~\ref{sec:noise}, we characterize the impact of decoherence on the SNAP gate by computing the state fidelity for preparing the Fock state \( |1\rangle \).  
An alternative approach is to directly evaluate the gate fidelity of the SNAP gate in the presence of noise.  
Following a procedure similar to the closed-system gate fidelity calculation in Sec.~\ref{sec:floquet_snap}, we first construct the reference propagator \( \hat{\mathcal{E}}_\text{ref} \), which describes evolution under the sideband drive alone.  
To achieve this, we solve a set of Lindblad master equations for various initial conditions \( |g,n;t=0\rangle \langle g,m;t=0| \), obtaining the results at $t=t_\text{g}$ \( |g,n;t=t_\text{g}\rangle \langle g,m;t=t_\text{g}| \), where \( m,n \in [0,N_\text{c}] \).  
Using vectorization, we extract the matrix elements of \( \hat{\mathcal{E}}_\text{ref} \):  
\begin{equation*}
    \hat{\mathcal{E}}_\text{ref}^{[(m,n), (m',n')]} = \langle g, (m,n); t=0 | g, (m',n'); t=t_\text{g}\rangle,
\end{equation*}
where \( | g, (m,n); t\rangle \) represents the vectorized form of \( |g,n;t\rangle \langle g,m;t| \).  
Similarly, we construct the propagator \( \hat{\mathcal{E}} \) when both the sideband and SNAP drives are applied.  
The target propagator \( \hat{\mathcal{E}}_\text{target} \) is derived from the closed-system propagator \( \hat{U}_\text{target} \) and transformed from the interaction picture to the lab frame as $\hat{\mathcal{E}}_\text{lab} = \hat{\mathcal{E}}_\text{ref}\hat{\mathcal{E}}_\text{target} $. 
The gate fidelity is then computed as $\text{Tr}(\hat{\mathcal{E}}_\text{lab}^\dag \hat{\mathcal{E}})/N_\text{c}^2$.
In Fig.~\ref{fig:FIG_APP_OPEN_FIDELITY}, we plot the open-system gate infidelity for the 1500-ns QOC Floquet SNAP gate, as implemented in Fig.~\ref{fig:FIG_QOC_FLOQUET_SNAP}.  
In particular, we present the fidelity as a function of the truncated Hilbert space dimension \( N_\text{c} \).  
Overall, the results show a weak dependence on \( N_\text{c} \) and closely match the state fidelity obtained in Sec.~\ref{sec:qoc_snap}, shown as the dashed line.

\nocite{data_availability}
\bibliography{main.bib}
\end{document}